\documentclass[preprint2]{aastex631}

\usepackage{xcolor}
\usepackage{CJKutf8}

\definecolor{myred}{HTML}{BF3465}

\def\ltsima{$\; \buildrel < \over \sim \;$}
\def\simlt{\lower.5ex\hbox{\ltsima}}
\def\gtsima{$\; \buildrel > \over \sim \;$}
\def\simgt{\lower.5ex\hbox{\gtsima}}

\newcommand {\um}{$\mu$m}

\newcommand {\msun}{M$_{\odot}$}

\newcommand {\Cii}{[C\textsc{ii}]}
\newcommand {\Oiii}{[O\textsc{iii}]}

\def\ltsima{$\; \buildrel < \over \sim \;$}
\def\simlt{\lower.5ex\hbox{\ltsima}}
\def\gtsima{$\; \buildrel > \over \sim \;$}
\def\simgt{\lower.5ex\hbox{\gtsima}}

\received{--}
\revised{--}
\accepted{--}
\submitjournal{ApJ}

\shorttitle{ASPIRE: A Quasar-Anchored Protocluster at z=6.6}
\shortauthors{Champagne et al.}

\begin{document}

\title{A Quasar-Anchored Protocluster at $z=6.6$ in the ASPIRE Survey: I. Properties of [OIII] Emitters in a 10 Mpc Overdensity Structure}

\correspondingauthor{Jaclyn B. Champagne}
\email{jbchampagne@arizona.edu}

\suppressAffiliations
\author[0000-0002-6184-9097]{Jaclyn~B.~Champagne}
\affiliation{Steward Observatory, University of Arizona, 933 N. Cherry Ave, Tucson, AZ 85721, USA}
\author[0000-0002-7633-431X]{Feige Wang}
\affiliation{Steward Observatory, University of Arizona, 933 N. Cherry Ave, Tucson, AZ 85721, USA}
\affiliation{Department of Astronomy, University of Michigan, 1085 S. University Ave., Ann Arbor, MI 48109, USA}
\author[0000-0002-4321-3538]{Haowen Zhang \begin{CJK}{UTF8}{gbsn}(张昊文)\end{CJK}}
\affiliation{Steward Observatory, University of Arizona, 933 N. Cherry Ave, Tucson, AZ 85721, USA}
\author[0000-0001-5287-4242]{Jinyi Yang}
\affiliation{Steward Observatory, University of Arizona, 933 N. Cherry Ave, Tucson, AZ 85721, USA}
\affiliation{Department of Astronomy, University of Michigan, 1085 S. University Ave., Ann Arbor, MI 48109, USA}
\author[0000-0003-3310-0131]{Xiaohui Fan}
\affiliation{Steward Observatory, University of Arizona, 933 N. Cherry Ave, Tucson, AZ 85721, USA}
\author[0000-0002-7054-4332]{Joseph F. Hennawi}
\affiliation{Department of Physics, University of California, Santa Barbara, CA 93106-9530, USA}
\author[0000-0002-4622-6617]{Fengwu Sun}
\affiliation{Steward Observatory, University of Arizona, 933 N. Cherry Ave, Tucson, AZ 85721, USA}

\author[0000-0002-2931-7824]{Eduardo Ba\~nados}
\affiliation{Max Planck Institut f\"ur Astronomie, K\"onigstuhl 17, D-69117, Heidelberg, Germany}
\author[0000-0001-8582-7012]{Sarah E. I. Bosman}
\affiliation{Institute for Theoretical Physics, Heidelberg University, Philosophenweg 12, D–69120, Heidelberg, Germany}
\affiliation{Max-Planck-Institut f\"{u}r Astronomie, K\"{o}nigstuhl 17, 69117 Heidelberg, Germany}
\author{Tiago Costa}
\affiliation{School of Mathematics, Statistics and Physics, Newcastle University, Newcastle upon Tyne, NE1 7RU, UK}
\author[0000-0003-2895-6218]{Anna-Christina Eilers}
\affiliation{Department of Physics, Massachusetts Institute of Technology, Cambridge, MA 02139, USA}
\affiliation{MIT Kavli Institute for Astrophysics and Space Research, Massachusetts Institute of Technology, Cambridge, MA 02139, USA}
\author{Ryan Endsley}
\affiliation{Department of Astronomy, University of Texas Austin, 2515 Speedway Blvd, Austin, TX 78712, USA}
\author[0000-0002-5768-738X]{Xiangyu Jin}
\affiliation{Steward Observatory, University of Arizona, 933 N. Cherry Ave, Tucson, AZ 85721, USA}
\author[0000-0003-1470-5901]{Hyunsung D. Jun}
\affiliation{Department of Physics, Northwestern College, 101 7th St SW, Orange City, IA 51041, USA}
\author[0000-0001-6251-649X]{Mingyu Li}
\affiliation{Department of Astronomy, Tsinghua University, Beijing 100084, China}
\author[0000-0001-6052-4234]{Xiaojing Lin}
\affiliation{Steward Observatory, University of Arizona, 933 N. Cherry Ave, Tucson, AZ 85721, USA}
\author[0000-0003-3762-7344]{Weizhe Liu}
\affiliation{Steward Observatory, University of Arizona, 933 N. Cherry Ave, Tucson, AZ 85721, USA}
\author[0000-0002-8857-6784]{Federica Loiacono}
\affiliation{INAF - Osservatorio di Astrofisica e Scienza dello Spazio di Bologna, via Gobetti 93/3, I-40129, Bologna, Italy}
\author[0000-0001-6106-7821]{Alessandro Lupi}
\affiliation{Dipartimento di Scienza e Alta Tecnologia, Universit\`a degli Studi dell'Insubria, via Valleggio 11, I-22100, Como, Italy}
\affiliation{INFN, Sezione di Milano-Bicocca, Piazza della Scienza 3, I-20126 Milano, Italy}
\affiliation{Dipartimento di Fisica ``G. Occhialini'', Universit\`a degli Studi di Milano-Bicocca, Piazza della Scienza 3, I-20126 Milano, Italy}
\author[0000-0002-5941-5214]{Chiara Mazzucchelli}
\affiliation{Instituto de Estudios Astrof\'{\i}sicos, Facultad de Ingenier\'{\i}a y Ciencias, Universidad Diego Portales, Avenida Ejercito Libertador 441, Santiago, Chile}
\author[0000-0003-4924-5941]{Maria Pudoka}
\affiliation{Steward Observatory, University of Arizona, 933 N. Cherry Ave, Tucson, AZ 85721, USA}
\author{Klaudia Protu\v{s}ov\`{a}}
\affiliation{Institute for Theoretical Physics, Heidelberg University, Philosophenweg 12, D–69120, Heidelberg, Germany}
\affiliation{Max-Planck-Institut f\"{u}r Astronomie, K\"{o}nigstuhl 17, 69117 Heidelberg, Germany}
\author[0000-0003-2349-9310]{Sof\'ia Rojas-Ruiz}\affiliation{Department of Physics and Astronomy, University of California, Los Angeles, 430 Portola Plaza, Los Angeles, CA 90095, USA}
\author[0000-0003-0747-1780]{Wei Leong Tee}
\affiliation{Steward Observatory, University of Arizona,
933 N. Cherry Ave, Tucson, AZ 85721, USA}
\author{Maxime Trebitsch}
\affiliation{Kapteyn Astronomical Institute, University of Groningen, P.O Box 800, 9700 AV Groningen, The Netherlands}
\author[0000-0001-9024-8322]{Bram P.\ Venemans}
\affiliation{Leiden Observatory, Leiden University, Einsteinweg 55, NL-2333 CC Leiden, the Netherlands}
\author[0000-0001-5105-2837]{Ming-Yang Zhuang 
\begin{CJK}{UTF8}{gbsn}(庄明阳)\end{CJK}}
\affiliation{Department of Astronomy, University of Illinois Urbana-Champaign, Urbana, IL 61801, USA}
\author[0000-0002-3983-6484]{Siwei Zou}
\affiliation{Chinese Academy of Sciences South America Center for Astronomy, National Astronomical Observatories, CAS, Beijing 100101, China}

\begin{abstract}

ASPIRE (A SPectroscopic survey of bIased halos in the Reionization Era) is a quasar legacy survey primarily using \textit{JWST} to target a sample of 25 $z>6$ quasars with NIRCam slitless spectroscopy and imaging.
The first study in this series found evidence of a strong overdensity of galaxies around J0305$-$3150, a luminous quasar at $z=6.61$, within a single NIRCam pointing obtained in JWST Cycle 1.
Here, we present the first results of a JWST Cycle 2 mosaic that covers 35 arcmin$^2$ with NIRCam imaging/WFSS of the same field to investigate the spatial extent of the putative protocluster.
The F356W grism data targets \Oiii+H$\beta$ at $5.3<z<7$ and reveals a population of 124 line emitters down to a flux limit of 1.2$\times$10$^{-18}$\,erg\,s$^{-1}$\,cm$^{-2}$.
Fifty-three of these galaxies lie at $6.5<z<6.8$ spanning 10 cMpc on the sky, corresponding to an overdensity within a 2500 cMpc$^3$ volume of 12.5 $\pm$ 2.6, anchored by the quasar. 
Comparing to the \Oiii\, luminosity function from the Emission line galaxies and Intergalactic Gas in the Epoch of Reionization (EIGER) project, we find a dearth of faint \Oiii\, emitters at log(L/erg\,s$^{-1}$) $<$ 42.3, which we suggest is consistent with either bursty star formation causing galaxies to scatter around the grism detection limit or modest suppression from quasar feedback.
While we find a strong filamentary overdensity of \Oiii\, emitters consistent with a protocluster, we suggest that we could be insensitive to a population of older, more massive Lyman-break galaxies with weak nebular emission on scales $>10$\,cMpc.

\end{abstract}


\section{Introduction} \label{sec:intro}

Dedicated surveys have now revealed a population of $>200$ rare, extremely bright quasars at $z>6$, many with measured black hole masses in excess of 10$^9$\,\msun\, \citep[e.g.,][]{Banados2016a, Yang2019a, Wang2021a, Fan2023a, Mazzucchelli2023a, Banados2023a}.
The existence of these billion-solar-mass black holes $<1$ Gyr after the Big Bang poses a major challenge to our understanding of supermassive black hole formation \citep[see review in][]{Volonteri2021a}. 
These bright $z>6$ quasars are broadly predicted by various simulations to reside in massive host galaxies within the rarest, most massive dark matter halos in the early Universe ($M_h \sim 10^{12.5-13}$ \msun{}; \citealt{Costa2014a,Angulo2012a,DiMatteo2017a, Lupi2024a}).
This is observationally corroborated by their strong clustering \citep{Garcia-Vergara2017a, Arita2023a}.

Given their expected halo masses, the earliest quasars should trace very strong matter overdensities associated with the seeds of present-day galaxy clusters \citep[though the variance on scales larger than a few cMpc is significant, e.g.,][]{Angulo2012a}.
Prior to \textit{JWST}, much effort was invested towards quantifying galaxy overdensities around $z>6$ quasars based on imaging using either \textit{HST} or large ground-based telescopes.
But while overdensities were occasionally found in observations, they certainly were not ubiquitous \citep[e.g.,][]{Kim2009a, Banados2013a, Mazzucchelli2017b, Ota2018a, Champagne2023a, RojasRuiz2024a}.
Interpreting these results was extremely challenging given heterogeneous selection techniques, restricted fields of view, and different cosmic volumes probed by various galaxy tracers (e.g., Lyman-break galaxies, Ly$\alpha$ emitters, and dusty star-forming galaxies).
Moreover, the limited near-IR sensitivity and poor photometric redshift precision ($\Delta z \gtrsim 1$) afforded by ground and \textit{HST} surveys likely caused all but the strongest true galaxy overdensities around $z>6$ quasars to be missed.
This was also coupled with the uncertainty of the underlying spatial distribution of companion halos, compounding the issue of observational geometry \citep[e.g.][]{Zana2023a}.

Thus, the question of whether the earliest quasars routinely trace strong galaxy overdensities could not be answered without near-IR spectroscopy in wide fields of view \citep[e.g.,][]{Lupi2022a}.
Dark matter simulations predict that galaxy protoclusters are extended beyond tens of comoving Mpc at $z>6$ \citep{Chiang2017a, Muldrew2015a, Overzier2016a}, which cannot be probed with, for example, single ALMA pointings \citep{Champagne2018a}.
Now, with the power of JWST/NIRCam's wide-field slitless spectroscopy (WFSS) in fields of view on the scale of tens of square arcminutes, we can perform detailed case studies around individual quasar environments.

To this end, two major Cycle 1 surveys emerged targeting the environments of reonization-era quasars. 
The Emission-line galaxies and Intergalactic Gas in the Epoch of Reionization (EIGER) project \citep{Kashino2023a, Matthee2023a, Eilers2024a} has already begun to use NIRCam's WFSS mode to characterize the environments and statistical clustering of a sample of six $z\sim6$ quasars, finding a diversity of overdensity signals.
The second, the focus of this paper, is A SPectroscopic survey of bIased halos in the Reionization Era (ASPIRE) project \citep[][]{Wang2023a, Yang2023a}.
Both of these studies perform grism spectroscopy in F356W with direct imaging in F356W, F200W, and F115W.

In total, ASPIRE targeted 25 $z>6.5$ quasars in \textit{JWST} Cycle 1 with single NIRCam pointings.
This paper focuses on a single quasar field from the ASPIRE sample, which is well-studied at multiple wavelengths.
J0305$-$3150 was originally identified in the VIKING Survey \citep{Venemans2013a, Venemans2016a} and lies at $z=6.61$, containing a $\sim10^9$\,\msun\, SMBH \citep{Mazzucchelli2017a}.
\citet{Ota2018a} found the first hints of an overdensity in this field, using Subaru broad- and narrowband imaging to identify 53 LBGs and 14 LAEs, corresponding to a 3$\sigma$ and 1$\sigma$ overdensity respectively across a 30\arcmin\, by 30\arcmin\, field.
Later, \citet{Champagne2023a} found an overdensity of LBGs in the field, with $\delta_{\rm gal} \equiv N_{\rm obs}/N_{\rm exp} -1 = 8.8 \pm 1.8$ based on photometric redshift fitting of 42 galaxies identified with \textit{HST} broadband imaging within a 6.25 arcmin$^2$ field of view (FOV).
Using ASPIRE Cycle 1 data, \citet{Wang2023a} identified 41 galaxies at $5.4<z<6.9$ in a single NIRCam pointing (11 arcmin$^2$) of the same field via the detection of \Oiii+$\rm H\beta$, 21 of which were within $\Delta z \pm 0.2$ ($\sim 7800$\,km\,s$^{-1}$) from the quasar.
Of those twenty-one, 13 \Oiii\, emitters were matched to 10 spatially-unresolved LBGs from \citet{Champagne2023a}.
\citet{Wang2023a} thus provided evidence for a spectroscopic overdensity extending several comoving Mpc on the sky, which motivated us to target this field again in Cycle 2 with a NIRCam mosaic covering 6$\times$ the area to investigate the extended protocluster structure.

With spectroscopic data in hand and a suite of imaging from prior studies \citep{Ota2018a, Champagne2023a} and NIRCam \citep{Wang2023a}, we can investigate in detail the rest-optical properties of these galaxies and compare them to the general field population towards the end of the epoch of reionization.
In this paper (Paper I), we present the galaxies detected in a NIRCam mosaic of the field of J0305$-$3150 and evaluate the protocluster nature of the overdensity.
The following paper \citep[Paper II;][]{Champagne2024b} presents detailed SED fitting of the galaxies using the full suite of imaging and discusses the environmental dependence of galaxy evolution within the protocluster.
We describe our dataset and reduction process with details on our catalog construction for \Oiii\, emitters and LBGs in \S\ref{sec:obs} and \S\ref{sec:catalog}. 
\S\ref{sec:protocluster} shows the 3D distribution of the filaments in the protocluster.
\S\ref{sec:o3lf} presents the \Oiii\, luminosity function and equivalent width distribution for the spectroscopically confirmed protocluster members.
In \S\ref{sec:discussion} we present our interpretation of the environment of the quasar within the protocluster, while we compare to simulations in \S\ref{sec:sims} and conclude in \S\ref{sec:conclusion}. 
Throughout this paper we assume AB magnitudes and a flat $\Lambda$CDM cosmology with H$_0$ = 70\,km\,s$^{-1}$\,Mpc$^{-1}$, $\Omega_{\Lambda}=0.7$, and $\Omega_{\rm M}$=0.3. 

\section{Observations and Data Reduction}\label{sec:obs}

\subsection{JWST Data}\label{sec:jwst}
J0305$-$3150 was observed as part of the Cycle 1 \textit{JWST} ASPIRE (A SPectroscopic survey of biased halos In the Reionization Era) program (GO \#2078, PI: F. Wang) which targets 25 $z>6.5$ quasars with F356W grism spectroscopy and F115W/F200W/F356W broadband imaging with NIRCam.
More details about the ASPIRE survey can be found in \citet{Wang2023a}. 
Follow-up mosaic observations were performed in Cycle 2 (GO \#3325, PI: F. Wang).
We use the ASPIRE grism spectroscopy  to identify \Oiii$\lambda$5007 emitters, which are complemented by the additional 5-pointing mosaic in Cycle 2 centered on the quasar, for a total area coverage of 35.05 arcmin$^2$.
We briefly summarize the data reduction and processing to homogenize the data here.

\subsubsection{NIRCam WFSS}
ASPIRE uses Grism-R with F356W in the long wavelength (LW) channel ($\rm R\sim1300-1600$), with simultaneous observations with F200W in the short wavelength channel (SW). 
Thus, the quasars are observed with slitless spectroscopy at 3-4\,\um\, with deep ($\sim$28 mag at 5$\sigma$) imaging at 2\,\um.
The main observations are performed with a 3-point \verb|INTRAMODULEX| primary dither pattern and each primary position includes two sub-pixel dithers, yielding a survey area of $\sim11$ arcmin$^2$ per pointing for imaging + slitless spectroscopy (the mosaicked area is 35 arcmin$^2$ accounting for overlaps).
We use the \verb|SHALLOW4| readout pattern with nine groups and one integration which gives a total on-source exposure time of 4257 s per pointing, with the deepest data centered on the quasar. 
Full details of the reduction steps including astrometric and spectral calibrations, dispersion modeling, and extraction of spectra can be found in \citet{Wang2023a}.
That study noted a half-pixel offset between the spectral tracing model and the data along the spatial direction, but that the offset along the dispersion direction requires in-flight wavelength calibration that is still not available.
They quote a conservative constant offset of $<$100\,km\,s$^{-1}$ which translates to $\Delta z < 0.003$ for the \Oiii\, emitters, which we adopted here as well.

\subsubsection{NIRCam imaging}
To maximize the sky area coverage, both NIRCam modules are used by ASPIRE for direct imaging.
In order to match spectra to their sources from the grism spectroscopy, direct (near the quasar) and out-of-field imaging (to capture sources outside of the NIRCam field of view whose spectra may land on the WFSS detector) were performed with the same readout pattern as the main observations, with the F115W filter in the SW and the F356W filter in the LW. 
Therefore, J0305 was observed with F115W, F200W, and F356W, with the latter being the deepest.
Reduction of the NIRCam images was performed using version 1.10.2 of the JWST Calibration Pipeline (\verb|CALWEBB|).
We use the reference files (\verb|jwst_1015.pmap|) from version 11.16.21 of the standard Calibration Reference Data System (CRDS) to calibrate our data. 
The details of the Stages 2 and 3 steps, including creating background images, the measurement of the \textit{1/f} noise, astrometric alignment, image drizzling and final background subtraction can be found in \citet{Wang2023a, Yang2023a}.
During the resampling step, we used a fixed pixel scale of 0.031\arcsec\, for the SW images and 0.0315\arcsec\, for the LW images with adopted \verb|pixfrac|=0.8. 
The mosaicked images are further aligned to the reference catalog from \textit{Gaia} DR3 \citep{Gaia2020a} 
for absolute astrometric calibration, yielding precise relative alignment (RMS $\sim$ 15 mas) and absolute astrometric calibration (RMS $\sim$ 50 mas).
The 5$\sigma$ depths, calculated by placing random empty 0\farcs32-diameter apertures across the image, are 27.2, 28.0, and 28.3 in F115W, F200W, and F356W respectively.

\section{Catalog Construction}\label{sec:catalog}

\begin{figure*}
    \centering
    \includegraphics[width=2.0\columnwidth]{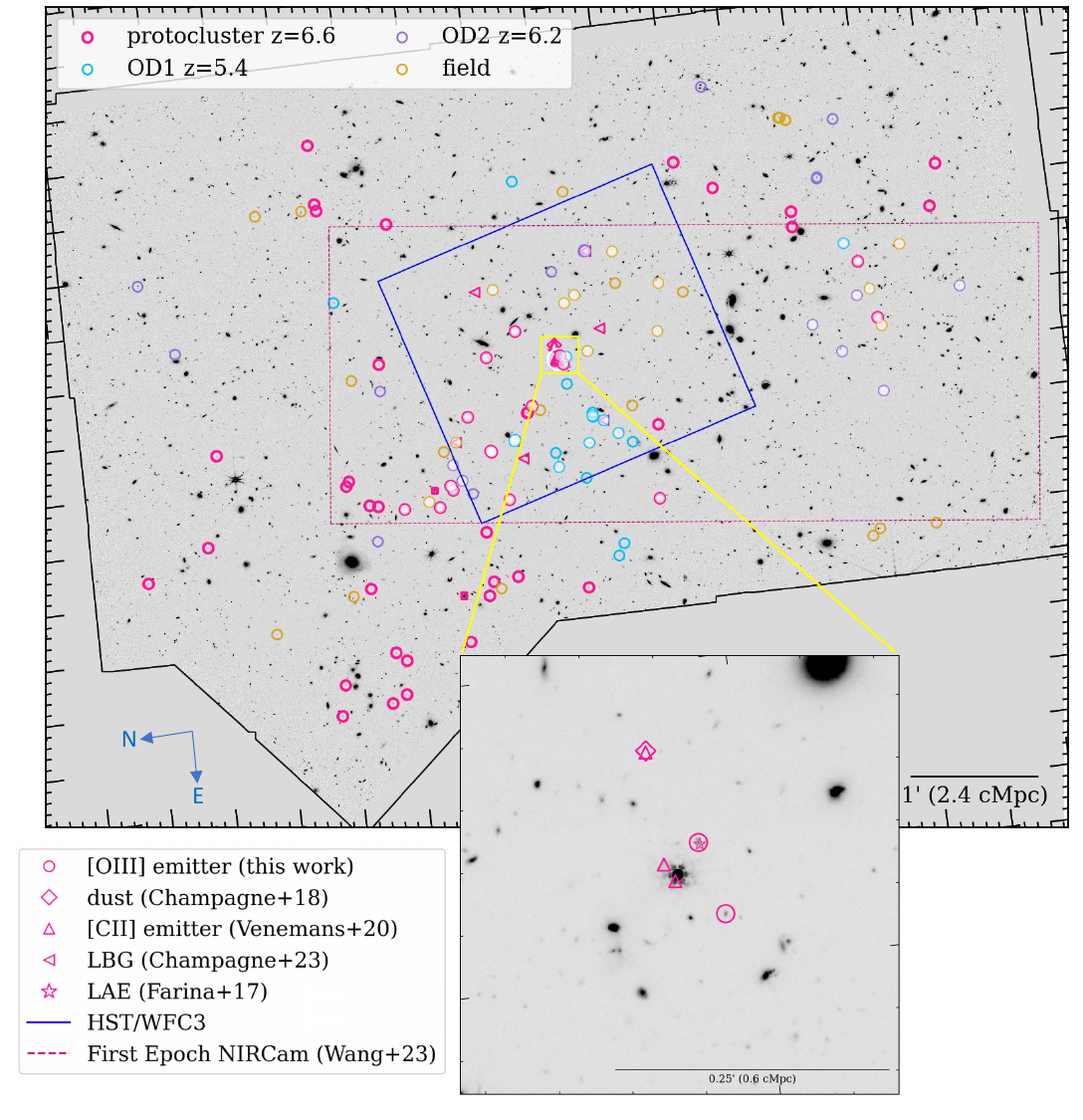}
   \caption{
    Full F356W six-pointing mosaic surrounding J0305$-$3150, with the field of view outlined in black. The quasar is depicted at the center with a white circle. The quasar-anchored overdensity members are shown in pink, the overdensities at $z=5.4$ and $z=6.2$ are in blue and purple, and the field galaxies at $5.3<z<6.5$ are shown in yellow. The blue box shows the WFC3 footprint where galaxies have photometric coverage from \textit{HST}, and the dashed plum line shows the single NIRCam pointing from \citet{Wang2023a} with the quasar centered in Module A. The configuration of the pointings is such that we have the greatest depth in the immediate surroundings of the quasar, so we are most sensitive to faint galaxies in the central regions. The filled points indicate sources already identified in the Cycle 1 data \citep{Wang2023a}.
  Inset: Zoom-in to the central 30\arcsec\, surrounding the quasar. Three companion \Cii\, emitters \citep{Venemans2020a}, 1 dust continuum emitter \citep{Champagne2018a}, 1 LAE \citep{Farina2017a}, and 2 \Oiii\, emitters (this work; one of them is the LAE) are visible within a few hundred ckpc from the quasar.}
    \label{fig:f356w}
\end{figure*}

\begin{figure}
\centering
\includegraphics[width=1.0\columnwidth]{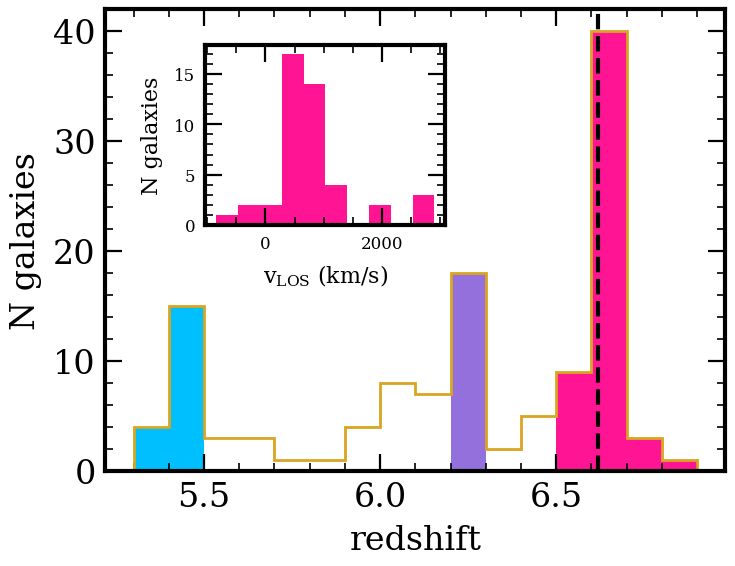}
\caption{Redshift histogram of all 124 \Oiii\, emitters identified here, with the three primary overdensities highlighted. The quasar is at $z=6.614$ \citep[dashed line;][]{Venemans2019a}. The inset shows the velocity distribution of galaxies with respect to the quasar, i.e. the galaxies hereafter considered to be members of the quasar-anchored protocluster.}
\label{fig:zhist}
\end{figure}

\subsection{The [OIII] catalog}\label{sec:o3cat}
We used \texttt{SourcExtractor++} \citep{Bertin2020a} on the NIRCam imaging to create an initial photometric catalog.
Details of the \texttt{SourcExtractor++} and photometric measurements can be found in Paper II; we use this catalog only as positional priors to extract \Oiii\, line emitter candidates.
We cross-matched our photometric catalog with the catalog of \Oiii\, emitters reported by \citet{Wang2023a}, which was extracted only from the central pointing of our mosaic data, using a search radius of 0\farcs1 (chosen to be close to the PSF size and to accommodate potential astrometric offsets from the Cycle 1 dataset). 
\citet{Wang2023a} reported 41 \Oiii\, emitters at $5.3<z<6.9$, all of which remain robust and with no positional offsets after using improved reduction procedures.

For the ASPIRE dataset, we \citep[][Wang et al, in prep.]{Wang2023a} developed a semi-automatic emission line candidate searching algorithm based the continuum-subtracted 2D and 1D spectra.
Briefly, for every source identified in the F356W image, we extract a grism spectrum using optimal extraction, which fits and weights a radial profile to the source, without concerns of flux loss as from a boxcar extraction.
We produce a median-filtered continuum model with a window size of 51 pixels, which gets subtracted from the extracted spectrum.
In a S/N spectrum smoothed to filter out hot pixels and artifacts, we search for S/N$>$1.2 peaks. 
To determine if a source is an \Oiii\, emitter candidate at $z\sim5-7$, we first assume any identified lines with $\rm S/N>5$ are from \Oiii$\lambda5008$ and then search for a corresponding \Oiii$\lambda4960$ and/or H$\beta$ line detected with SNR$>$2.
Sources with three securely identified lines are retained as high-quality candidates, though about 50\% of our sources have upper limits (S/N$<4$) on H$\beta$ fluxes.
An example spectrum is displayed in Appendix \ref{sec:appA}.

Combining with the new 5-pointing mosaic across 35 arcmin$^2$, our updated line-search procedure yielded 83 new \Oiii\, emitters which are added to our sample.
This corresponds to a total of 53 \Oiii\, emitters that are members of the overdensity at $z\geq6.5$ and 71 at $z<6.5$ (124 in all).
The 5$\sigma$ limiting flux in the deepest central part of the mosaic, assuming a linewidth of 50\,\AA\, (2$\times$ the spectral resolution, approximately 250\,km\,s$^{-1}$), is 1.2$\times$10$^{-18}$\,erg\,s$^{-1}$\,cm$^{-2}$.
This corresponds to a luminosity of $6\times10^{41}$ erg\,s$^{-1}$ at $z=6.6$. 
There are two serendipitously discovered overdensities along the line of sight at $z=5.4$ and $z=6.2$ which we will discuss in detail in \S\ref{sec:o3lf}.

Rest-frame equivalent widths (EW) are measured for the \Oiii\, emitters using the \Oiii\, line fluxes (obtained by integrating the best-fit Gaussian to the \Oiii\, doublet in the 1D spectrum) and the F356W broadband photometry after subtracting out the line contribution, assuming a flat continuum in $f_{\nu}$.

\subsection{The LBG catalog}\label{sec:lbgcat}
We use the catalog from \citet{Champagne2023a} who identified LBGs using a single pointing of 5-band \textit{HST} WFC3 and ACS imaging.
We do not identify any new LBGs within the \textit{HST} footprint that are not already in the \Oiii\, catalog, nor do we add any new \textit{HST} data covering the WFSS mosaic.
\citet{Champagne2023a} searched for LBGs in a wide redshift range of $\Delta z = 1.5$ due to the coarse sampling of the SED with only \textit{HST}.
We cross-matched that LBG catalog with our new photometric catalog which used F356W as the detection image (see Paper II) and included the NIRCam photometry to refit their SEDs using \texttt{EAZY} \citep{Brammer2008a}.
Of the 42 $z=5.9-7.6$ LBGs reported in \citet{Champagne2023a}, three are not detected in the F356W catalog, but they were classified as ``marginal" in the original work, corresponding to low-SNR ($<4.5\sigma$) detections in F160W and $<$80\% of the photometric redshift PDF lying at $z>6$; therefore, we discard them as likely not to be high-redshift galaxy candidates given the deeper imaging with NIRCam.
Cutouts of these discarded LBGs are shown in Appendix \ref{sec:appB}.
Of the remaining 39 LBGs, 18 do not have photo-$z$'s consistent with $z_{\rm qso}\pm0.3$ after the inclusion of NIRCam and Subaru photometry (this is the median offset between $z_{\rm \Oiii}$ and $z_{\rm phot}$; see paper II for template SED fitting), 11 are confirmed \Oiii\, emitters not at the quasar redshift ($5.4<z<6.3$), and 3 \Oiii\, emitters are already included in our primary overdensity sample.
The remaining 7 are still robust LBGs with $z_{\rm phot}\geq 6.31$.

Since the grism data is sensitive to a wide range of $5.3<z<7$, any LBGs from \citet{Champagne2023a} should have been detected in \Oiii\, even for those with high photo-$z$ uncertainty.
While they might be lower-$z$ interlopers given the sparse rest-UV/optical filter coverage, the total lack of emission lines in the remaining LBGs could be due to an inherent faintness of nebular emission, which could be caused by rapid bursts and lulls in the star formation history occurring on $<$10 Myr timescales \citep[e.g.,][]{ Endsley2024a, Faisst2024a, Trussler2025a}.
We hypothesize that they have lower specific star formation rates which would result in undetectable \Oiii\, emission at the ASPIRE flux limit, a consequence of stochastic star formation histories (discussed further in Paper II).
For those LBGs which are not confirmed in \Oiii\, but have robust photo-$z$ estimates (defined as $>80$\% of the EAzY redshift PDF lying between $6.4<z<6.8$), we derive an upper limit on the EW using the observed F356W magnitude and a 5$\sigma$ limiting line flux at the location of the LBGs ($\approx 2\times10^{-18}$\,erg\,s$^{-1}$\,cm$^{-2}$). 

\section{A Protocluster at $z=6.61$}\label{sec:protocluster}
The quasar overdensity consists of 53 galaxies at $6.5<z<6.8$ ($\Delta V \approx$\,10,000\,km\,s$^{-1}$)\footnote{This is quite a bit larger than the typical protocluster definition used in the literature of 1000$-$2000\,km\,s$^{-1}$ along the line of sight, but because the overdensity's redshift distribution is a roughly continuous Gaussian (Figure \ref{fig:zhist}), we consider all galaxies in this velocity range to be members for now.}, with a maximum extent of 417\arcsec\, or 17 proper Mpc.
Within the overdensity, a surprising 41 lie at exactly the quasar redshift ($\Delta V < 1000$\,km\,s$^{-1}$). 
Two other overdensities are revealed in the full distribution of \Oiii\, emitters: a compact overdensity at $z=5.35-5.45$ ($\Delta V \approx$\, 4000\,km\,s$^{-1}$) consisting of 20 galaxies extended across 240\arcsec\, (10 pMpc), and a more extended overdensity (385\arcsec\, or 16 pMpc) at $z=6.2-6.3$ ($\Delta V \approx$\, 3000\,km\,s$^{-1}$) composed of 18 galaxies.
The remaining 33 line emitters between $5.5<z<6.2$ comprise our field sample. 
This serendipitous confluence of line-of-sight overdensities helps to explain the high number of LBGs found in this field by \citet{Champagne2023a}, but the number of galaxies with \Oiii-based spectroscopic confirmation at the quasar redshift remains remarkable.

\begin{figure}
\centering
\includegraphics[width=1.0\columnwidth]{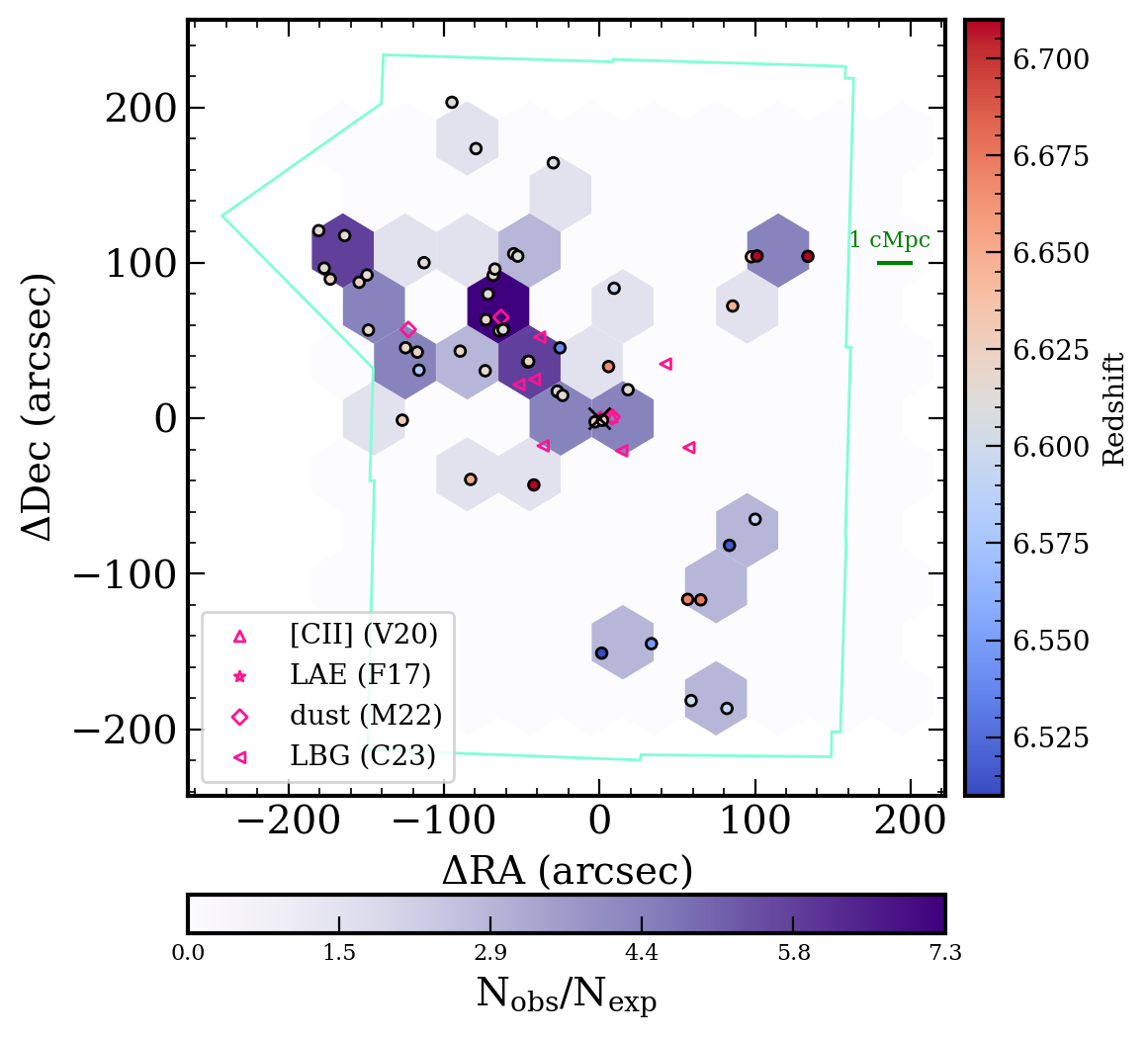}
\caption{2D representation of spectroscopically confirmed \Oiii\, emitters at the quasar redshift (circles). The quasar is marked as the large black $\times$; galaxies identified via other tracers are marked in pink symbols. The hexbin histogram (bottom colorbar) denotes the 2D overdensity of \Oiii\, emitters with respect to the UV luminosity function at $z=6$ \citep{Finkelstein2015a}, showing that the overdensity is stronger in a filament pointing away from the quasar. The right colorbar denotes the redshift of individual galaxies, centered on the quasar redshift. The green outline depicts the footprint of the mosaic.}
\label{fig:hexbin}
\end{figure}

Figure \ref{fig:f356w} shows the F356W image with all of the \Oiii\, emitters overlaid, denoting the quasar overdensity members, the lower redshift overdensities, and the field galaxies, in addition to other galaxies (dust continuum and \Cii) identified in the field by ALMA \citep{Champagne2018a, Venemans2020a, Meyer2022a}.
The redshift histogram of all 124 line emitters is shown in Figure \ref{fig:zhist}, highlighting the three primary overdensities.
The inset shows all galaxies within $\Delta z \pm 0.15$ from the quasar; while these galaxies are likely not \textit{all} associated with the same structure at such a wide line-of-sight distance, the overdensity signal remains strong across a continuous range in redshift at $6.5<z<6.8$ and a transverse area of (10 cMpc)$^2$, so we consider them all to be members for the sake of this study.
Basic information about the \Oiii\ emitters in the protocluster can be found in Table \ref{table:o3pc}.

Figure \ref{fig:hexbin} shows a 2D representation of the overdensity around the quasar, highlighting a filament extending behind the quasar across a transverse distance of $\sim5$\,cMpc.
The 2D overdensity hexbins are calculated based on the area expected to contain 1 LBG at $M_{UV} < -19$ according to the luminosity function at $z=6.6$ parameterized by \citet{Finkelstein2016a}.
The quasar actually does not sit at the center of the overdensity, but instead inhabits the SW side of the spatial distribution.
Further, it lies at the lower end of the galaxy redshift distribution by about 500\,km\,s$^{-1}$.
The full overdensity extends well into the NIRCam mosaic, with galaxies at the quasar redshift found within a (10 cMpc)$^2$ box on the sky.
In fact, the overdensity may extend well beyond the current FOV given that many sources are found on the comparatively shallow edges of the mosaic (see exposure map in Figure \ref{fig:exposure}).
It is distinctly distributed across multiple overdense filaments, with the densest region found about $1-2$ cMpc from the quasar at very slightly lower redshift ($z_{\rm qso} = 6.614$, $z_{\rm fil} = 6.618$).
This is comparable to the redshift uncertainty of $\delta z = 0.003$ and could potentially be due to peculiar motion, but regardless the quasar still does not lie at the spatial center of the overdensity.
Observationally, protoclusters identified at lower redshift in the fields of DSFGs or AGN are often not centered on the ``main" galaxy \citep[e.g.,][]{Dannerbauer2014a, Cucciati2018a, Toshikawa2024a} and are observed in an unrelaxed (i.e., non-spherical) distribution.
From a theoretical perspective, the BlueTides simulation \citep{DiMatteo2017a} finds that the most massive SMBH are not necessarily in the most spatially overdense regions, but instead in specific environments that favor radial matter inflows (perhaps in this case, on the edge of the overdensity).

Notably, as seen in the inset of Figure \ref{fig:f356w} and in Figure \ref{fig:hexbin}, the immediate environment of the quasar is characterized by a rich population of neighboring galaxies.
There are multiple submm-detected galaxies \citep{Venemans2020a} not seen in \Oiii, and there is a relative dearth of \Oiii\, emitters compared to the strong overdensities further from the quasar; we revisit the physics of this in \S\ref{sec:discussion}.
This lack of line emitters in the immediate vicinity of the quasar has been seen in other studies tracing LAEs, albeit with much larger ``holes" on the scale of 5 pMpc \citep[e.g.,][]{Lambert2024a}. 
Given the richness of the overdensity within a relatively small area on the sky --- consistent with theoretical expectations \citep[e.g.,][]{Overzier2009a} --- we strongly suspect that this is a bona fide galaxy protocluster, i.e. a progenitor to a massive cluster of M$_h \sim 10^{14}-10^{15}$\,\msun\, at later times \citep[e.g.,][]{Costa2014a}.

\section{The \Oiii\, Sample}\label{sec:o3lf}
\subsection{Luminosity function}
We next construct the luminosity function (LF) of our \Oiii\, emitters and compare our findings with the blank field \Oiii\, LF measured in the $5.3<z<6.9$ in the fore/background of quasars from the EIGER project \citep{Matthee2023a, Kashino2023a}.
The luminosity function is the usual formula:
\begin{equation}
\phi [\rm Mpc^{-3} dlogL] = \frac{N}{V_{max} C}
\end{equation}
where $C$ is the completeness, $N$ is the number of objects per bin, and $V_{\rm max}$ is the detectable volume subtended by our survey.
We bin the \Oiii-emitters in bins of log($L$/erg s$^{-1}$) = 0.2 and compute V$_{\rm max}$ as a function of redshift and luminosity.
Since the sensitivity of NIRCam WFSS is position- and wavelength-dependent, this translates to a redshift dependence of the detectability of \Oiii\, in addition to the overall luminosity limit. 
The survey volume is computed following \citet{Sun2023a}: for a specific redshift, we compute the effective sky area based on the spectral tracing and grism dispersion models in order to construct RMS maps, done using continuum-subtracted WFSS \texttt{stage2 cal} files. 
The maximum sky area at each redshift is the area of the RMS map with values smaller than the maximum RMS for a line detection of log($L$/erg s$^{-1}$) = 42.

\begin{figure}
\centering
\includegraphics[width=0.95\columnwidth]{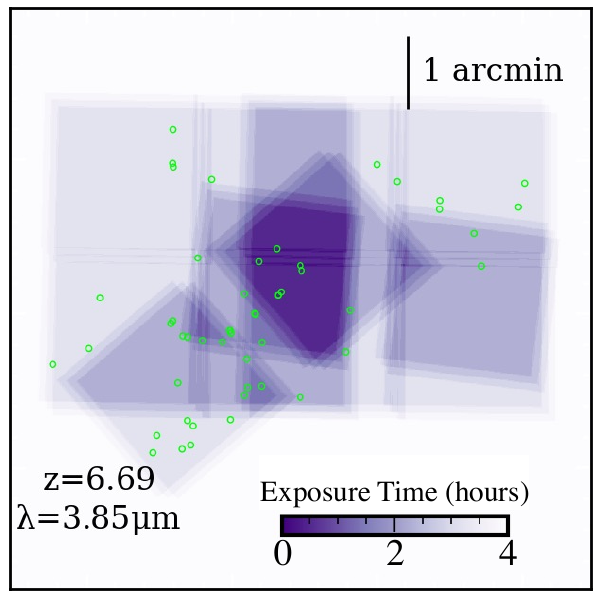}
\includegraphics[width=1.0\columnwidth]{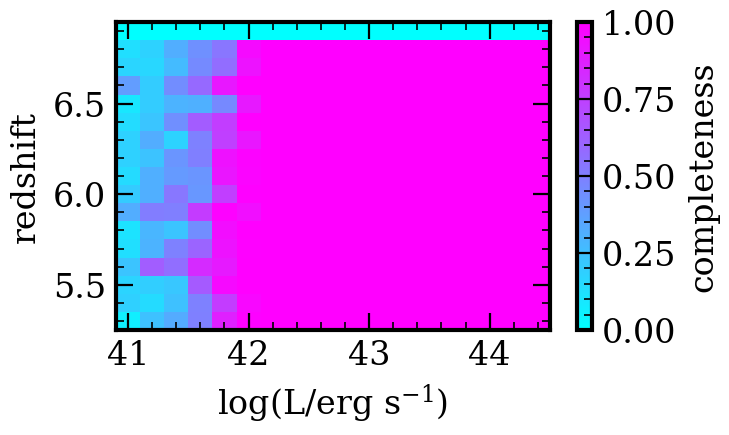}
\caption{\textit{Top:} Exposure map for the F356W grism mosaic centered at $\lambda_{\rm obs}$=3.85\,$\mu$m, or $z=6.69$. The locations of the protocluster members are denoted with green circles. \textit{Bottom:} Results of our 1D completeness simulations as a function of redshift and input line luminosity, calculated for the deepest central part of the mosaic. The completeness at each location is scaled by the measured RMS of a given line relative to the RMS in the deepest part of the mosaic.}
\label{fig:exposure}
\end{figure}

To measure completeness, we run 1D source injection simulations.
We begin with a noise spectrum with the average line-free rms of our sample at the deepest point of the mosaic (centered on the quasar) which is perturbed within 1$\sigma$ for every realization.
Then we insert Gaussian emission lines with the same wavelength resolution as the real data at a randomly sampled range of intrinsic luminosities ($41<\rm log(L/erg s^{-1})<44$), redshifts ($5.3<z<7$), and intrinsic FWHM (drawing from a Gaussian centered on 200 km/s with $\sigma=50$ km/s).
We then re-fit the Gaussian to the line and measure the recovered flux and SNR.
The completeness varies as a function of wavelength, so we repeat this procedure in bins of $\delta z = 0.1$ such that the completeness is measured as a function of both luminosity and redshift.
Figure \ref{fig:exposure} shows the exposure map for the mosaic as well as the results of our completeness simulations in the deepest part of the mosaic.
We find a 100\% recovery rate of galaxies at the quasar redshift above log(L/erg s$^{-1}$) = 42.0 in the center of the mosaic at $z=6.69$.
For each source at a given location on the exposure map, the value of the completeness moves along the luminosity axis by the ratio of the RMS at that location relative to that in the deepest exposure\footnote{Note that NIRCam Module A is about 20\% more sensitive than Module B in the same exposure time, but this is accounted for by using the measured RMS at a given location in the mosaic.
}.

\begin{figure*}
\centering
\includegraphics[width=1.0\columnwidth]{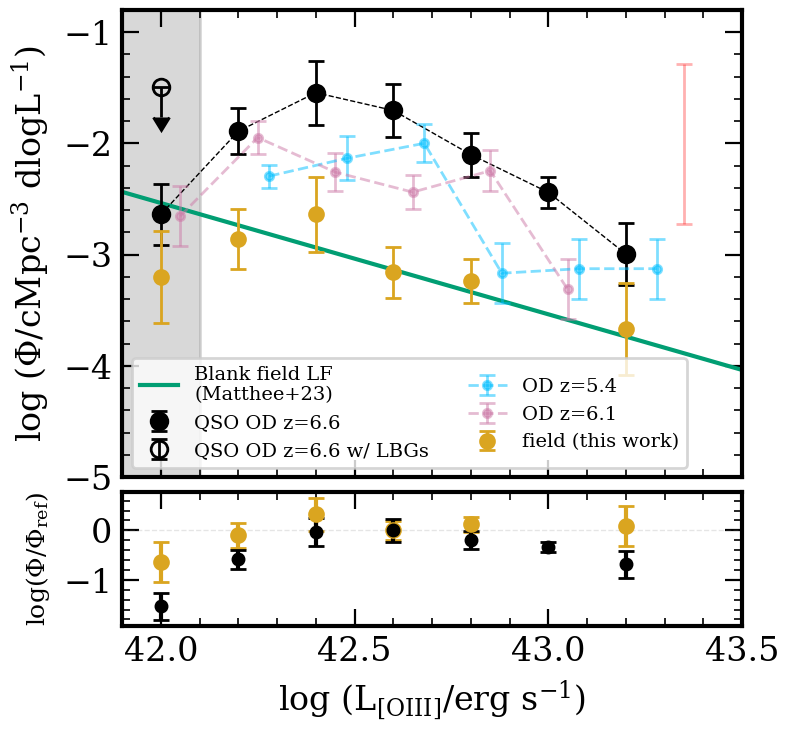}
\includegraphics[width=1.0\columnwidth]{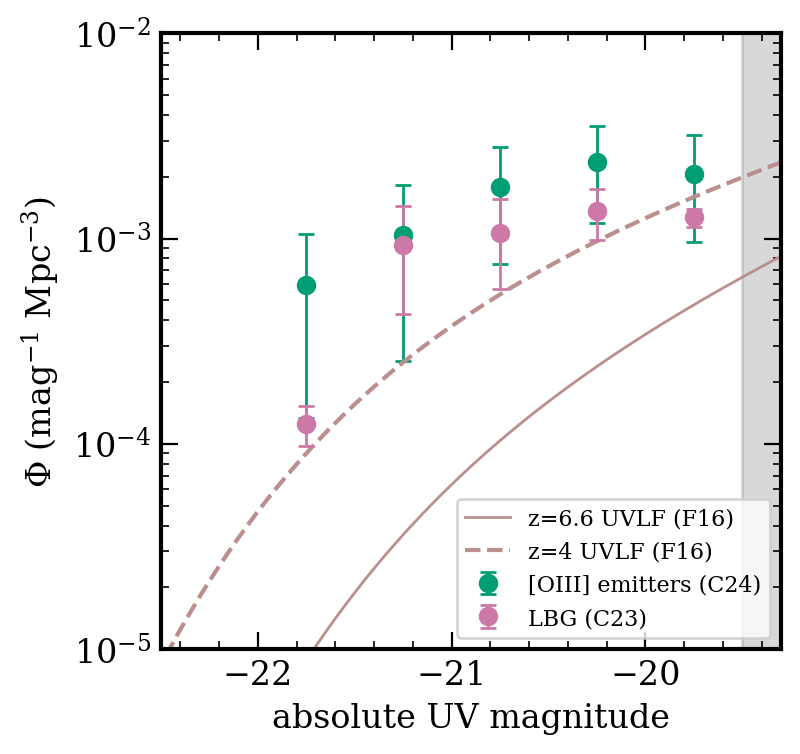}
\caption{\textit{Left top:} Luminosity function of the \Oiii\, emitters in our sample. The black points indicate members of the protocluster (empty point indicates the upper limit when including LBGs not detected in \Oiii); the blue and pink points indicate the overdensities at $z<6.5$. The green line is the blank field luminosity function from EIGER \citep{Matthee2023a}, and the filled gold points are the field galaxies in our sample that do not belong to the three overdensities. The grey shaded regions in both panels indicate where we are $<80$\% complete, and the red error bar indicates the characteristic error attributed to cosmic variance. The quasar overdensity is an order of magnitude above the blank field expectations but the shape is dissimilar, with a preference for fainter galaxies and a slight dearth of the brightest galaxies. \textit{Left bottom:} The ratio of the field (gold) and protocluster (black) luminosity functions to the EIGER blank field LF, normalized at log(L/erg\,s)=42.6, showing that the protocluster turnover is significant at $>3\sigma$ while the field sample is consistent with EIGER within 2$\sigma$ at any luminosity. \textit{Right:} The UV luminosity function within the quasar protocluster of our \Oiii\, emitters and LBGs from \citet{Champagne2023a}. M$_{UV}$ is calculated from the best-fit SEDs presented in Paper II. The redshift-parameterized UVLFs at $z=6.6$ and $z=4$ from \citet{Finkelstein2016a} are shown: both the \Oiii\, emitters and the LBGs are consistent with having undergone more rapid evolution than the field by several hundred Myr.}
\label{fig:o3lf}
\end{figure*}

Figure \ref{fig:o3lf} shows the results of our \Oiii\, luminosity function for the overdensity samples and the field sample.
We calculate the quasar overdensity factor by first integrating the \Oiii\, luminosity function derived by \citet{Matthee2023a} for the EIGER project, who searched for \Oiii\, emitters with an identical observational setup to ASPIRE in the field of the $z=6.3$ quasar J0100+2802 in a 26\,arcmin$^2$ mosaic.
We normalize the EIGER LF by the area (35\,arcmin$^2$) and volume covered by our field to calculate the expected number of \Oiii\, emitters between $6.5<z<6.8$ and arrive at a lower limit of $\delta_{\rm gal} \equiv N_{\rm obs}/N_{\rm exp} - 1$ = 3.7 $\pm$ 1.5.
However, the 53 galaxies in the overdensity are distinctly clustered in a (10 cMpc)$^2$ box on the sky (roughly 200$\times$200 arcsec$^2$), so if we instead integrate the EIGER luminosity function in this smaller region to arrive at the expected number of \Oiii\, emitters, $\delta_{\rm gal} = 12.5\, \pm$ 2.6. 
This is consistent with $\delta_{\rm gal} = 12$ found by \citet{Wang2023a} in a single NIRCam pointing of the J0305 field.
Thus, the filamentary structure extending well into the NIRCam mosaic footprint remains $>10\times$ overdense with respect to the field at least out to R=10 cMpc (possibly further, given that the structure extends to the edge of the mosaic), consistent with the size of protoclusters at $z>6$ \citep[e.g.,][]{Chiang2017a}.

The luminosity function of the quasar overdensity reveals two curious features: the slope of the protocluster LF at the bright end exhibits a sharper decline than the field relation, and there is a turnover in the protocluster LF at log($L$/erg s$^{-1}) = $ 42.4.
By scaling the blank field luminosity function in the foreground of the EIGER quasar J0100+2802 to the area covered by our survey, we would have expected to find a
handful of bright (log($L$/erg s$^{-1}) > $ 43) objects (14 $\pm$ 3 in the field at $5.3<z<6.5$, 2.2 $\pm$ 1 in the quasar overdensity), but this is not reflected either in the field nor the protocluster.
A simple Kolmogorov-Smirnov (KS) test between the protocluster \Oiii\, flux distribution and the field (normalized to the comoving volume of the protocluster within $\Delta z = 0.3$) shows a $p$-value of 0.35, so the two distributions are not distinguishable.
However, we are limited by comparing only two fields that are likely subject to strong field-to-field variance \citep[see, e.g., more results from EIGER;][]{Eilers2024a}. The lack of bright galaxies is not statistically significant, as cosmic variance is particularly dominant at the bright end, contributing an additional 78\% to the fractional error on the number counts \citep{Trenti2008a}.

More interesting than the lack of \Oiii-bright galaxies, however, is the turnover at the faint end. 
Our 5$\sigma$ limiting line luminosity at $z=6.6 $ is $6\times10^{41}$ erg/s, and we are presumably complete above $10^{42}$ erg/s at all redshifts between $5.4<z<6.8$ according to our 1D injection simulations.
Further, the field is statistically consistent with no turnover, showing agreement with the EIGER field LF within 1$\sigma$ at log(L/erg s$^{-1}$)$<$42.2, so we do not believe this is due to completeness. 
If we normalize the observed protocluster LF to the expected number in the field at log(L/erg $s^{-1}$)=42, we find that the two faintest bins are discrepant with the field at 3$\sigma$ significance.
Therefore, we explore the possibility that this turnover is due to a physical scenario affecting the observed luminosity distribution rather than a selection effect.

A genuine dearth of faint galaxies compared to bright galaxies within the protocluster could point to physics governing the strength of \Oiii\, emission, which varies on the timescale of the lifetime of O stars \citep[about 10 Myr;][]{Eldridge2022a}. 
A population of intrinsically faint (i.e. below the grism detection limit) galaxies with respect to their nebular emission could point to the idea that the majority of the galaxies in the protocluster are undergoing less recent star formation.
This could be the result of more massive and evolved galaxies having higher metallicity and higher continuum, and thus weaker \Oiii.
This is supported by the fact that, if we assume the 7 LBGs are at the quasar redshift and calculate the upper limits of their \Oiii\, luminosity by using their observed F356W magnitude and assuming the median EW of our sample, they would be below our nominal detection completeness limit but could contribute to the faintest bin in our LF --- effectively, the turnover in the luminosity function disappears (empty circle in the left panel of Figure \ref{fig:o3lf}).
However, we cannot immediately rule out that quasar feedback could also influence the number counts of faint \Oiii\, emitters. 
Yet we also note, qualitatively, that a similar downturn at the faint end of the LF is seen in the two lower-redshift overdensities.
This could imply that it is not the mere existence of the quasar influencing the \Oiii\, distribution (though, normalized to the EIGER LF at log(L/erg\,s$^{-1}$)=42 as above, the discrepancy in the lower-redshift overdensities is significant only at the 2--3$\sigma$ level).
We return to this point in Section \S\ref{sec:sims}.

Figure \ref{fig:o3lf} also shows the luminosity function of the two lower-redshift overdensities mentioned in \S\ref{sec:o3cat}. 
These show a similar order of magnitude overdensity above the EIGER relationship and our field LF.
It is beyond the scope of this paper to declare whether these are also protocluster structures, but we note it would be interesting to follow them up with further observations.
An analysis of the galaxy properties of all three overdensities (quasar and line-of-sight) is left to Paper II.

\begin{figure}
\centering
\includegraphics[width=1.0\columnwidth]{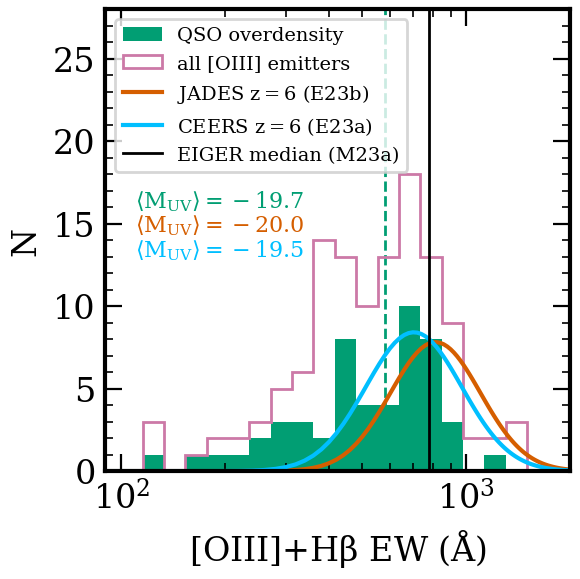}
\caption{Distribution of \Oiii+H$\beta$ equivalent widths in the full sample of 124 emitters (pink) and the protocluster sample (green, with median value shown with a dashed line). We compare to the JADES ``bright" sample (orange) from \citet{Endsley2023b} and the CEERS ``faint" sample (blue) from \citet{Endsley2023b}, with the median M$_{UV}$ of each sample labeled in the corresponding colors. The median EW from EIGER \citep{Matthee2023a} is shown in black. This sample, despite having similar rest frame UV magnitudes and redshift, trends towards weaker line emission than previous studies of field galaxies.}
\label{fig:ewdist}
\end{figure}

\subsection{\Oiii+H$\beta$ strength}
Next, in Figure \ref{fig:ewdist} we turn to the nebular line equivalent width distribution, which is a proxy for the ratio of young stellar populations (nebular emission lines powered by OB stars) to older populations (powering the stellar continuum).
The median EW of the protocluster population measured from the grism spectra is 580 $\pm$ 15\,\AA. 
The sample of non-overdensity galaxies in the field shows a marginally higher median and spread with 630$\pm$130\,\AA; this is more comparable to the median EW of 650$^{+110}_{-90}$\,\AA\, of UV-bright galaxies in COSMOS \citep{Endsley2021a, Whitler2023a}.
On the other hand, \citet{Endsley2023a} finds that the median \Oiii+H$\beta$ EW is 780 \AA\, for UV-faint (M$_{UV}\sim-19.6$) LBGs at $6<z<8$ with a dispersion of $\sim0.3$ dex; an even higher median of 890\,\AA\, is found for the JADES sample at similar redshift \citep{Endsley2023b}.
However, we note that \citet{Endsley2023b} and \citet{Endsley2021a} derive their EW from photometric excesses rather than spectra, which have photometric redshift uncertainties and could induce overestimates of the EW if the continuum or contribution 
of other faint lines are underestimated. 
To instead compare with spectroscopic samples, our measured median EW is also below the EIGER median stack value of 845 $\pm$ 70\,\AA\, \citep{Matthee2023a}, discrepant with our protocluster at 3$\sigma$.
This could be a systematic effect in our data, but we note that the EIGER median stack excludes galaxies at the redshift of the quasar, so this could potentially suggest a physical difference in the nebular line emission between galaxies within and outside of overdensities --- deeper data or observations of more quasar fields would be required to draw more firm conclusions.

Studies differ on the definition of an extreme emission line galaxy (EELG) but here we use the definition of EW(\Oiii+H$\beta$) $>$ 750\AA\, \citep[e.g.,][]{Boyett2024a}.
Only 10/53 (19 $\pm$ 10\%) of the protocluster galaxies satisfy this definition while 11/33 (33 $\pm$ 15\%) field galaxies can be considered EELGs.
Thus, the fraction of EELGs in the protocluster is lower at the 2$\sigma$ level according to Poisson and small number statistics.
It is consistent with the fact that, despite our area coverage whichis wide enough to pick up a handful of rare bright objects, we observe the majority of the \Oiii\, emitters in the protocluster to be rather faint, pointing to an environmental effect.
We therefore suggest that the EW distribution in the protocluster could be biased low due to the presence of more evolved galaxies with earlier formation times, resulting in low nebular emission and high continuum from the dominating presence of moderately-aged stars. 
While we acknowledge that the field ($N=33$) and protocluster ($N=53$) samples are consistent with being drawn from the same EW distribution (with a KS $p$-value of 0.21) and that the EELG fraction is discrepant only at 2$\sigma$, the discrepancy between the protocluster and much larger field samples from blank field surveys remains strong at $>3\sigma$.

Going beyond the spectroscopic data, the question remains, are quasar companion galaxies preferentially brighter compared to the field? 
We finally compare our results to the UV luminosity function derived in the field \citep{Finkelstein2016a}.
\citet{Champagne2023a} suggested a preferential enhancement of bright ($M_{UV}<-20$) LBGs in the vicinity of $z>6$ quasars, including J0305$-$3150, even after correcting the faint end for completeness.
Computing the LBG luminosity function in similar bins as the field UV luminosity function of \citet{Finkelstein2016a}, they found that the slope and normalization of the LBG luminosity function in the field of J0305$-$3150 was most consistent with the field UVLF at $z=4$, suggestive of accelerated evolution compared to the field.
Now that the \Oiii\, emitters in the same field are spectroscopically confirmed, we can return to the question of whether brighter galaxies prevail in the overdensity, or galaxy evolution is enhanced in the full mass range we probe.

While the \Oiii\, luminosity distribution peaks at relatively low luminosities, we find that the $M_{\rm UV}$ distribution of the \Oiii\, emitters (measured from their rest-frame SEDs at 1500 \AA, see Paper II for SED details) follows the same slope as that of the LBGs in \citet{Champagne2023a}: a flattening of the bright end most consistent with evolution accelerated by several hundred Myr (Figure \ref{fig:o3lf}, right panel).
This is in line with descendant populations, i.e., more evolved bona-fide protoclusters, at $z=2-4$.
For example, \citet{Hill2020a} and \citet{Pensabene2024a} and  both find preferential enhancements of bright galaxies within protoclusters at cosmic noon, implying that more massive galaxies evolve faster than their lower-mass counterparts.
This is consistent with the distribution of low \Oiii+H$\beta$ EWs, implying older ages (and thus lower specific star formation rates) and higher metallicity among massive galaxies, as both effects contribute to lowering the H$\beta$+\Oiii\, equivalent width.

\section{Discussion}
\subsection{Immediate Quasar Environment}\label{sec:discussion}
Figure \ref{fig:f356w} clearly shows a very busy immediate ($\sim$100 kpc) environment around the quasar, marked by 3 \Cii\, emitters \citep{Venemans2019a}, a dust continuum emitter \citep{Champagne2018a}, one LAE \citep{Farina2017a}, and 2 \Oiii\, emitters from this work (one of which is the LAE)\footnote{In Paper II, we search for AGN activity in the protocluster members, but for now we assume all of the \Oiii\, emitters are normal star-forming galaxies.}.
Note that the three \Cii-emitters from \citet{Venemans2019a} and the dust continuum emitter from \citet{Champagne2018a} are each found in single ALMA pointings ($\sim25$\arcsec\, diameter in Band 6), but no other \Cii\, or dust sources are found in a 1.1\arcmin\, mosaic centered on the quasar (Wang et al., in prep.).

It is interesting to note the relative dearth of \Oiii-emitting galaxies within 100\,kpc from the quasar host compared to the much stronger overdensities found further away from the quasar (Figure \ref{fig:hexbin}).
One way to explain this could be that the quasar host galaxy has grown through major mergers with its immediate neighbors.
The \Cii\, emitters without \Oiii\, counterparts could imply the existence of massive gas reservoirs but with relatively low instantaneous star formation, 
which would result in a non-detection of the short-lived \Oiii\, line.
Indeed, \citet{Venemans2019a} suggest that the small molecular gas masses ($\sim5\times$ smaller than the quasar host) and unusual kinematics of the \Cii\, companion galaxies could imply prior interactions with the host galaxy.

A second explanation for the perceived lack of galaxies very close to the quasar is the result of high dust extinction in the rest-optical.
While Wang et al. (in prep.) finds no dust continuum counterparts for the \Oiii\, emitters nor new sources, the quantity of dust could be below the ALMA detection limit but still significant enough to weaken the \Oiii\, or indeed the Ly$\alpha$ line.
In fact, \citet{Ota2018a} imaged this field with on a much wider FOV with Subaru and found 14 narrowband-selected LAE candidates at the quasar redshift, but none closer than 2 arcmin (0.5 cMpc) from the quasar. 
Galaxies very close to the quasar could thus be undergoing a dustier mode of star formation than those on the outskirts of the overdensity.
\cite{Venemans2019a} measures SFRs for the \Cii\, companions in the range of 25--160\,\msun\,yr$^{-1}$ --- which, again, have no \Oiii\, counterparts --- so indeed some vigorous star formation is taking place not accounted for by our \Oiii\, selection method.

A final way to explain the relative dearth of \Oiii\, emitters very close to the quasar host is through radiative feedback from the quasar.
Indeed, \citet{Yang2023a} present blueshifted \Oiii\, outflows from quasar hosts at similar redshift which could be the result of radiatively driven AGN feedback in kinetic mode, i.e., where radiation pressure produces outward motion of heated gas.
Photoionization heating from the central quasar can suppress star formation in surrounding low-mass haloes within the so-called proximity zone, whose size is largely determined by the quasar lifetime and UV luminosity \citep{Satyavolu2023a}.
The photoionization heating can be described by the quantity $J_{21}$ \citep{Kashikawa2007a, Bosman2020a}.
This quantity, relating the quasar's UV intensity at the Lyman limit (912\,\AA) to its environment, is typically applied to the low-mass haloes hosting Lyman-$\alpha$ emitters (LAEs).
Assuming that the low-mass \Oiii\, emitters (median stellar mass $10^8$\,\msun; see SED details in Paper II) occupy similar haloes, we calculate the UV flux density as a function of distance to the quasar using the following equation:

\begin{equation}
F_{\nu}^Q = \frac{L(\nu_L)}{4\pi r^2}
\end{equation}

where

\begin{equation}
L(\nu_L) = 3631 * 4\pi D_L^2 10^{-0.4m_{1450}} \left(\frac{912}{1450}\right)^{-\beta}
\end{equation}

where $m_{1450}$ = 20.89 mag \citep{Venemans2013a}, $\beta$ is the continuum slope which is measured to be $-0.66$ using archival photometry of the quasar (Protu\v{s}ov\`{a} in prep.), $r$ is the angular diameter distance from the quasar, and $D_L$ is the luminosity distance to the quasar.
Then $J_{21}$ is the isotropic UV intensity $J/10^{-21}$ erg\,s$^{-1}$\,cm$^{-2}$\,Hz$^{-1}$\,sr$^{-1}$ where $J=F_{\nu}^Q/4\pi$.
Evaluating these equations at 2 pMpc (15.2 cMpc, approximately the full angular extent of the protocluster) we find that $F_{\nu}^Q$ = 1.1$\times$10$^{-19}$ erg\,s$^{-1}$\,cm$^{-2}$\,Hz$^{-1}$, and thus $J_{21}$ = 9.8 $\pm$ 0.9.

Some studies \citep[e.g.,][]{Kashikawa2007a, Chen2020a} have suggested that values of $J_{21} > 1$ can completely suppress star formation in low-mass halos, 
but we do not see such an effect in our field as there are indeed at least 56 galaxies within 2 pMpc of the quasar.
For one, we expect that the \Oiii\, emitters occupy halos of $\sim10^{10}$\,\msun (see \S\ref{sec:sims}) which is above the halo mass limit where we expect to see significant suppression \citep[e.g.,][]{Bosman2020a}.
Secondly, not all galaxies in the vicinity of the quasar will be equally exposed to the quasar radiation due to 1) patchy dust obscuration within the host galaxy and 2) the fact that the UV flux from the quasar is radiated in a beam with a modest opening angle rather than isotropically. 
Such a double-cone shape is indeed accommodated by the non-spherical distribution of \Oiii\, emitters (Figure \ref{fig:hexbin}), though we note the mosaic is not uniformly sensitive across the whole area.
Modest suppression of slightly higher-mass halos may still be occurring in the very inner regions close to the quasar, where J$_{21} > 100$, given the low number of close companions.
This is supported by the shape of the \Oiii\, luminosity function (Figure \ref{fig:o3lf}), where we see potential hints of suppression of \Oiii\, emitters at the faint end (i.e., where the LF is the steepest.)
We explore in the next section whether the turnover in the LF could primarily be attributed to intrinsic star formation properties of the galaxies or a direct result of interaction with the quasar.

\subsection{Comparison to simulations}\label{sec:sims}
Given the extended filamentary nature of the quasar protocluster spanning over 12 cMpc on the sky and $\Delta z = 0.2$ along the line of sight, we can already conclude the structure is not virialized.
Simulations suggest that galaxies in the crowded cores of protoclusters will merge to become the brightest cluster galaxy (BCG) at early times \citep[$z>3$;][]{Rennehan2020a}, but the spatially-wide, non-spherical distribution of galaxies in this reionization-era protocluster does not imply that it is currently approaching any relaxed state.
Therefore, we cannot use the virial theorem to estimate a total halo mass for this structure, nor can we comment on whether this structure will ever fully collapse into a galaxy cluster.
Bearing this in mind, we can still evaluate the protocluster in the context of very large halos in the early Universe by considering halos with similar masses to that of the quasar host galaxy.
We finally wish to investigate whether the observed population of \Oiii-emitters follows what we would expect from simulations of such massive halos.

\begin{figure*}
 \centering
  \includegraphics[width=2.0\columnwidth]{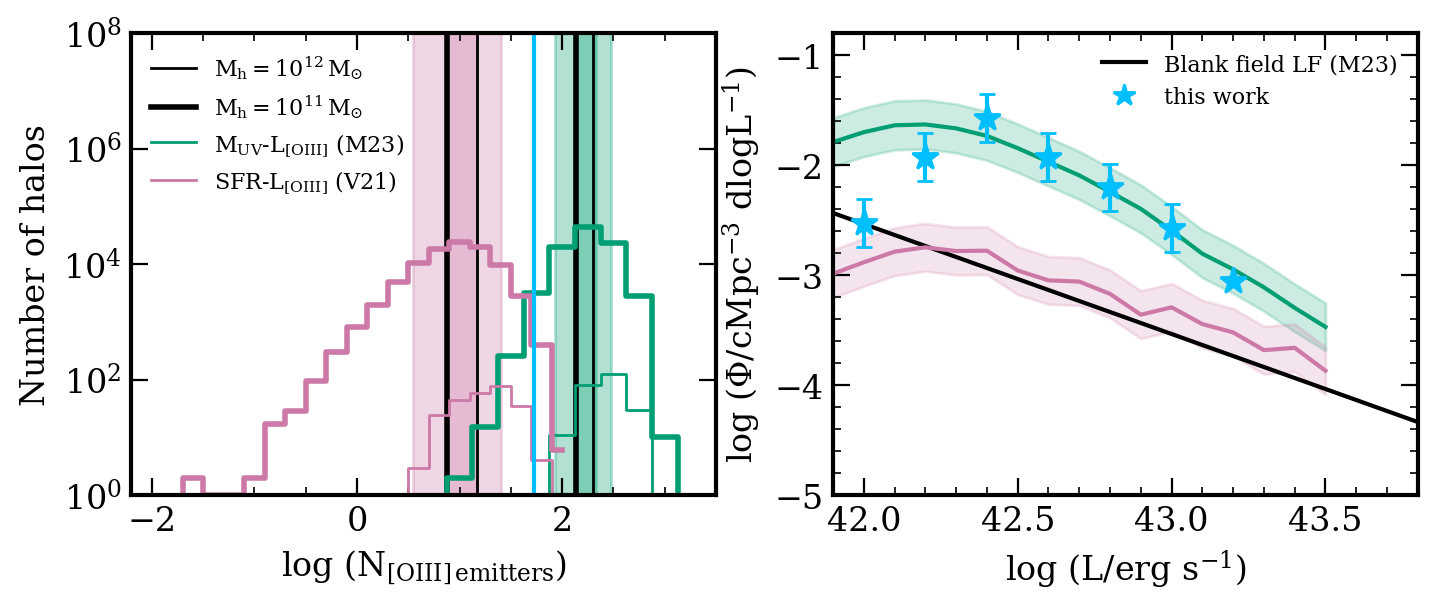}
   \caption{Predictions of \Oiii-emitting companions in $M_h>10^{11}$\,\msun\, halos extracted from \textsc{UniverseMachine} \citep{Behroozi2019a}. \textit{Left:} Number of galaxies predicted in our completeness-corrected footprint, calculated by assigning L$_{\Oiii}$ based on the predicted $M_{UV}$ \citep[green;][]{Matthee2023a} or SFR \citep[pink;][]{VillaVelez2021a}, with the thick and thin histograms denoting $M_h>10^{11}$\,\msun\ and $M_h>10^{12}$\,\msun\ halos respectively. The light blue line indicates the observed number of quasar overdensity members. The black solid lines are the median predicted number counts with the shaded regions corresponding to the 1$\sigma$ spread for that halo mass and \Oiii\, scaling relation. \textit{Right:} Simulation-predicted \Oiii\, luminosity functions, with the same color scheme and 1$\sigma$ errors as the left, compared to the observed LF (blue stars) and the EIGER LF (black line). The EIGER $M_{UV}-L_{[OIII]}$ relationship best predicts the observed shape in the quasar overdensity, but predicts a higher absolute number of companions at the faint end. This could be due either to suppression from the quasar or intrinsic properties of the star formation within the galaxies.}
    \label{fig:simulation}
\end{figure*}

Cosmological hydrodynamical simulations such as \textsc{BlueTides} \citep{DiMatteo2017a} as well as empirical models such as \textsc{Trinity} \citep{Zhang2023d} suggest that quasars hosting SMBH masses similar to J0305$-$3150 should occupy massive halos on the order of $M_h\sim10^{12}$\,\msun.
We first compare with halos and galaxies in the \textsc{UniverseMachine} mock catalogs based on the \textit{Small MultiDark-Planck} N-body simulation (\textit{SMDPL}, box size: 400 Mpc/$h$, particle number: 3840$^3$, halo mass limit: $\sim 10^{10} M_\odot$\footnote{This halo mass limit roughly corresponds to an \Oiii\, line flux of 10$^{42}$ erg/s, which is our 80\% completeness limit. The contribution from low-mass haloes to our measured companion number counts is expected to be small due to decreasing completeness below this limit.}; see \citealt{Klypin2016a}) as well as the quasar host halos based on the cosmological zoom-ins from \citet{Costa2024a}.

\subsubsection{UniverseMachine}
We use the \textsc{UniverseMachine} \citep{Behroozi2019a} to investigate the properties of galaxies in overdensities within massive halos.
Since \textsc{UniverseMachine} does not contain SMBH information and \textsc{Trinity} does not contain star formation histories (and thus ages) of individual galaxies, we use both together.
We use the empirical \textsc{Trinity} model \citep{Zhang2023a} to estimate a distribution of halo masses that could host a quasar with the luminosity and SMBH mass of J0305$-$3150.
To extract halos from \textsc{UniverseMachine}, we use the grism RMS footprint over the whole mosaic.
We convert the RMS to a 5$\sigma$ limiting line flux map by integrating a point source across a 250 km/s linewidth at $z=6.6$.
We extract 187,368 halos with $M_h > 10^{11}$\,\msun\, containing overdensities of galaxies in order to assess whether the quasar halo hosts an unusual distribution of galaxies compared to other massive halos in the field.
\textsc{UniverseMachine} provides, among other quantitites, $M_{UV}$ and star formation histories of the simulated galaxies populating each halo.

We assign \Oiii\, luminosities to the extracted galaxies using two methods: one sampling the (dust-uncorrected) SFR-L$_{\Oiii}$ relation from \citet{VillaVelez2021a} and one using the $M_{UV}-L_{\Oiii}$ relation from EIGER \citep{Matthee2023a}.
We find that the two methods result in a wide spread of predicted \Oiii\, emitters in our footprint, with the former method predicting $7-25$ companions with log($L/$erg\,s$^{-1}) > 42$ in a $M_h\sim10^{12}$\,\msun\ halo: a factor of a few below what we observe, but consistent with the blank-field luminosity function from EIGER (Figure \ref{fig:simulation}).
The latter method instead predicts $85-200$ companions compared to our observed 53, which again opens the question of suppressed SFR in galaxies exposed to quasar radiation.
The significant discrepancy between the predictions of these two methods likely has to do with the large scatter in the relationships between $M_{UV}$, \Oiii\, line luminosity, and instantaneous SFR as is being revealed by bursty star formation histories in both observations \citep{Endsley2024b} and simulations \citep[e.g., FLARES;][]{Wilkins2023a}.
Since UniverseMachine was calibrated on pre-JWST measurements of lower redshift galaxies with more continuous star formation, it is likely that its time resolution does not characterize the real stochasticity in these relationships.

Figure \ref{fig:simulation} shows the median theoretical \Oiii\, luminosity function for both methods mentioned above.
We plot the LF for galaxies residing in $M_h\sim10^{12}$\,\msun\, halos after applying the same $V_{\rm max}$ and completeness corrections as we did to our real data. 
The shape of the completeness-corrected theoretical overdensity LF measured using the EIGER $M_{UV}$ relation closely matches what we observe in the quasar overdensity, that is, a peak in the overdensity around log($L/$erg\,s$^{-1}) \sim 42.3$ and a faint-end turnover. 
However, the turnover is weaker than what we observe.
Modulo small number statistics, we might have expected only a handful more \Oiii-bright (log (L/erg s$^{-1}) >$ 43) galaxies within the quasar overdensity, so the overall paucity of observed \Oiii\, emitters compared to the simulation comes from the faint end.

On one hand, we posit that the grism data could be missing a real population of galaxies that are bright in the UV but have intrinsically weaker nebular emission \citep[or patchy dust content obscuring the emission, e.g.,][]{Faisst2024a} and scatter below our grism detection limit.
On the other hand, there is substantial scatter in the scaling relations used to predict \Oiii\, luminosity from the simulated observables, which will particularly impact the steep faint-end of the LF. 
Indeed, \textsc{UniverseMachine} does not capture the variance of star formation on timescales traced by \Oiii\, due to the coarse time resolution; thus it would not capture galaxies bright in the UV but with a recent downturn of star formation resulting in weak \Oiii\, emission, i.e. a departure from the canonical $M_{\rm UV}-$ or $SFR-L_{\rm \Oiii}$ relationships.

\subsubsection{Comparing to Costa (2024)}
Distinctly, \textsc{UniverseMachine} does not contain black hole physics, so our predictions do not encode any potential baryonic effects induced by the quasars in the center of the simulated fields.
We thus also compare to the simulated quasar halos from \citet{Costa2024a} who used cosmological zoom-in simulations based on the \texttt{Millennium} dark matter-only simulation, assuming steady-state ISM and star formation physics \citep{Springel2003a} and the $L_{\rm \Oiii}-M_h$ relation from \citet{Matthee2023a}.
\citet{Costa2024a} considered low-mass (`LM', $0.6-1\times10^{12}$\,\msun) and high-mass (`HM', $6-7\times10^{12}$\,\msun) quasar host halos and predicted the number counts of companions with moderate ($M_* = 1-10\times10^9$\,\msun) and high ($M_* > 10^{10}$\,\msun) mass. 
We can potentially rule out J0305 residing in a HM halo as, even within the variance, that study would predict tens of $M_*>10^9$\,\msun\, and 1-10 $M_*>10^{10}$\,\msun\, companions, while we find only 9 and 0 respectively (see paper II for calculation of stellar masses); this rejection is also supported by \textsc{Trinity}'s prediction of a host halo mass of $\sim1.1\times10^{12}$\,\msun. 
Yet \citet{Costa2024a} still predicts $1-10$ high-mass and $4-25$ moderate-mass companions even for the LM halos, while the majority of the galaxies in our sample have relatively low stellar mass (median log\,$M_*/M_{\odot} = 8.2$; see Paper II).
However, the simulation does not account for strong AGN feedback due to super-Eddington bursts and may not accurately model the dust attenuation, both of which could result in an overprediction of the number of companions.
On the other hand, this also suggests that we could be missing more massive galaxies with weak nebular line emission (therefore undetectable by grism), backed up by an observed anti-correlation between stellar mass and \Oiii+H$\beta$ EW \citep{Begley2025a}.
In this case, those galaxies can only be revealed with deeper and wider imaging bracketing the Lyman break and rest-UV/optical continuum, or ALMA observations targeting FIR lines. Alternatively, we could be still missing some physics governing the evolution of galaxies within overdensities. 
Indeed, the discovery of massive \Cii\, companions \citep{Decarli2018a, Venemans2019a} without \Oiii\, counterparts as well as seven LBGs consistent with the quasar redshift supports this argument. 

Overall, the faintness of the \Oiii\, emitters and the low-mass nature of the companions in the quasar overdensity paints a picture wherein galaxies experience stochastic episodes of star formation, causing galaxies to scatter in and out of the \Oiii\, detection limit on the scale of 5--10 Myr \citep[][]{Wilkins2023a, Faisst2024a}.
The fact that both simulations we compare to here predict a higher population of detectable bright and massive galaxies than what is observed is puzzling, but could be explained if we are not sensitive to massive galaxies with older stars and less \Oiii\, emission \citep[e.g.,][]{Looser2024a, Looser2023b}.
On the other hand, suppression of lower-mass halos by the central quasar engine could affect the presence of faint \Oiii\, emitters close to the quasar, as discussed previously.
While \textsc{UniverseMachine} does not resolve halos below $M_h\sim10^{10}$\,\msun\, where suppression should be strongest, the disagreement at the faint end between the observations and the simulation could be ascribed to the lack of strong quasar feedback in the simulation.

In the end, we are dealing with small number statistics in a relatively small area, and the simulations we compare to indeed predict a high amount of variance in the number of companions. 
The simulated \Oiii\, luminosities are highly sensitive to the chosen scaling relation since they are not directly predicted by the simulations, thus it is difficult to predict the absolute number of companions.
A larger sample of quasars, both hosting and not hosting overdensities, is required to answer these questions, which will indeed be delivered by the full ASPIRE sample (Wang et al., in prep.; Champagne et al., in prep.)

\section{Conclusions}\label{sec:conclusion}

In this work we have characterized a sample of 124 \Oiii\, emitters identified in the ASPIRE grism+imaging quasar legacy survey, identified over an area of 35 arcmin$^2$ surrounding the $z=6.6$ quasar J0305$-$3150.
53 of these galaxies are members of a quasar-anchored protocluster at $z=6.6$, while 18 and 20 galaxies occupy serendipitously discovered overdensities at $z=6.2$ and $z=5.4$. 
The remaining galaxies served as a field sample as a comparison between the evolution of galaxies within and without overdensities during the epoch of reionization.
We found that:

\begin{itemize}

\item The protocluster structure suggested by in \citet{Wang2023a} across 3 Mpc in a single NIRCam pointing in fact extends over 10 Mpc within a 35 arcmin$^2$ mosaic. The 53 galaxies at $6.5<z<6.8$ represent an overdensity of $\delta_{\rm gal} = 12.5 \pm 2.6$. They are distinctly distributed along filaments extending down the line of sight from the quasar.

\item The \Oiii\, luminosity function in the quasar protocluster has a distinct peak at log(L/erg s$^{-1}$) = 42.3, with a faint-end turnover that is not due to incompleteness. 
We suggest that we are missing a population of galaxies with \Oiii\, emission just below the grism detection limit since bursty star formation will affect the strength of \Oiii\, emission on very short (tens of Myr) timescales. 

\item There is a dearth of \Oiii\, emitters within 80\,ckpc from the quasar while there are three massive \Cii\, emitters and an LAE; we suggest that this could be due to mergers with the host galaxy, dust extinction, and/or mild suppression of low-mass \Oiii\, emitters due to photoionization heating from the quasar.

\item After computing the quasar host halo mass from \textsc{Trinity}, we compare our \Oiii\, flux distribution with galaxies in similar halos extracted from \textsc{UniverseMachine} and a hydrodynamic zoom-in simulation from \citet{Costa2024a}. The number of companion galaxies is very sensitive to the scaling relation used to paint on \Oiii\, luminosity, but our results are consistent with the protocluster overdensity being weaker than expected. This supports our argument that galaxies with stochastic SFHs are not picked up by this survey and would appear as LBGs with no line emission in a dedicated photometric survey. Indeed, a subset of our NIRCam data is covered by \textit{HST} which shows 7 LBGs without \Oiii\, counterparts but with photometric redshifts consistent with the protocluster; further multiwavelength imaging would be required across the full mosaic field of view to confirm this trend at larger distances.

\end{itemize}

A larger quasar sample will be presented in future ASPIRE publications which will help build up population statistics for quasar environments.
Wang et al. (in prep.) will present the ALMA counterpart to ASPIRE which includes mosaicked 1.1 arcmin $\times$1.1 arcmin Band 6 data in this field targeting dust continuum and \Cii.
In future studies, it will be crucial to obtain better photometric coverage of the rest-optical and NIR emission of many ASPIRE fields, especially through NIRCam medium bands and/or MIRI broadband photometry to probe low-sSFR galaxies not detectable by the grism spectroscopy.
It will also be interesting to compare the population of \Oiii\, emitters with a follow-up sample of Ly$\alpha$ emitters in the same field to assess the correlations between Ly$\alpha$ enhancement and high nebular line EW.
Paper II \citep{Champagne2024b} presents detailed SED fitting and an environmental analysis of the \Oiii\, emitters identified in this field.

\begin{acknowledgments}

JBC  acknowledges funding from the JWST Arizona/Steward Postdoc in Early galaxies and Reionization (JASPER) Scholar contract at the University of Arizona. 
FW and JBC acknowledge support from NSF Grant AST-2308258.
CM acknowledges support from Fondecyt Iniciacion grant 11240336  and ANID BASAL project FB210003.
FL acknowledges support from the INAF GO 2022 grant ``The birth of the giants: JWST sheds light on the build-up of quasars at cosmic dawn" and from the INAF 2023 mini-grant ``Exploiting the powerful capabilities of JWST/NIRSpec to unveil the distant Universe."
AL acknowledges support by the PRIN MUR ``2022935STW."
SZ acknowledges support from the National Science Foundation of China (grant no.\ 12303011).
SEIB is supported by the Deutsche Forschungsgemeinschaft (DFG) under Emmy Noether grant number BO 5771/1-1.
MT acknowledges support from the NWO grant 016.VIDI.189.162 ("ODIN").

This work is based on observations made with the NASA/ESA/CSA James Webb Space Telescope. The data were obtained from the Mikulski Archive for Space Telescopes at the Space Telescope Science Institute, which is operated by the Association of Universities for Research in Astronomy, Inc., under NASA contract NAS 5-03127 for JWST. These observations are associated with programs \#2078 and \#3225. Support for these programs was given through a grant from the Space Telescope Science Institute, which is operated by the Association of Universities for Research in Astronomy, Inc., under NASA contract NAS 5-03127.
The specific observations analyzed can be accessed via \dataset[10.17909/vt74-kd84]{https://doi.org/10.17909/vt74-kd84}.
\end{acknowledgments}

\software{Astropy \citep{astropy:2013, astropy:2018, astropy:2022}, SourceXtractor++ \citep{Bertin2020a}}

\begin{deluxetable*}{ccccccc}[h!]
\tablehead{\colhead{Name} & \colhead{RA} & \colhead{Dec} & \colhead{$z$} & \colhead{log L$_{\rm [OIII]}$} & \colhead{EW$_{\rm [OIII]}$} & \colhead{EW$_{\rm [OIII]+H\beta}$}\\ \colhead{} & \colhead{(J2000)} & \colhead{(J2000)} & \colhead{} & \colhead{10$^{42}$ erg s$^{-1}$} & \colhead{$\mathring{A}$} & \colhead{$\mathring{A}$}}
\tablecaption{\label{table:o3pc} Observational properties of \Oiii\, emitters in the protocluster. The sources are in the order they appear in the photometric detection catalog, with J0305-O3E-PC standing for J0305M3150--\Oiii\, emitter--protocluster member.
The \Oiii-derived redshift is assumed to have an uncertainty of $\delta z = 0.003$. All equivalent widths are in the rest frame, using the line-subtracted F356W photometry as the continuum.}
\startdata
J0305-O3E-PC-001 & 03:05:28.955 & -31:48:55.22 & 6.616 & $9.33\pm0.92$ & $423\pm50$ & $649\pm54$ \\
J0305-O3E-PC-002 & 03:05:28.724 & -31:49:19.32 & 6.615 & $6.66\pm0.77$ & $310\pm98$ & $444\pm103$ \\
J0305-O3E-PC-003 & 03:05:28.463 & -31:49:26.41 & 6.621 & $6.90\pm0.66$ & $452\pm224$ & $737\pm247$ \\
J0305-O3E-PC-004 & 03:05:27.846 & -31:48:58.42 & 6.616 & $10.96\pm0.86$ & $406\pm87$ & $681\pm96$ \\
J0305-O3E-PC-005 & 03:05:27.226 & -31:49:28.47 & 6.625 & $2.06\pm0.81$ & $119\pm35$ & $192\pm36$ \\
J0305-O3E-PC-006 & 03:05:26.887 & -31:49:23.75 & 6.614 & $3.85\pm0.70$ & $326\pm123$ & $425\pm128$ \\
J0305-O3E-PC-007 & 03:05:26.828 & -31:49:59.19 & 6.618 & $2.88\pm0.93$ & $198\pm24$ & $378\pm26$ \\
J0305-O3E-PC-008 & 03:05:25.235 & -31:50:10.56 & 6.621 & $3.18\pm0.88$ & $123\pm34$ & $301\pm37$ \\
J0305-O3E-PC-009 & 03:05:24.442 & -31:49:15.83 & 6.614 & $1.84\pm0.79$ & $518\pm166$ & $907\pm191$ \\
J0305-O3E-PC-010 & 03:05:25.373 & -31:50:57.16 & 6.628 & $18.14\pm0.91$ & $411\pm50$ & $639\pm55$ \\
J0305-O3E-PC-011 & 03:05:23.246 & -31:47:32.68 & 6.611 & $4.70\pm0.67$ & $498\pm195$ & $616\pm202$ \\
J0305-O3E-PC-012 & 03:05:24.732 & -31:50:13.34 & 6.623 & $3.53\pm0.78$ & $350\pm123$ & $617\pm135$ \\
J0305-O3E-PC-013 & 03:05:24.657 & -31:50:25.01 & 6.579 & $1.39\pm0.86$ & $577\pm97$ & $742\pm103$ \\
J0305-O3E-PC-014 & 03:05:22.216 & -31:48:02.45 & 6.609 & $2.32\pm0.83$ & $238\pm58$ & $361\pm60$ \\
J0305-O3E-PC-015 & 03:05:22.893 & -31:50:12.74 & 6.622 & $1.10\pm0.83$ & $197\pm44$ & $258\pm45$ \\
J0305-O3E-PC-016 & 03:05:21.698 & -31:49:36.02 & 6.615 & $3.46\pm1.00$ & $479\pm7$ & $745\pm12$ \\
J0305-O3E-PC-017 & 03:05:21.478 & -31:49:23.91 & 6.617 & $3.71\pm0.90$ & $349\pm50$ & $504\pm53$ \\
J0305-O3E-PC-018 & 03:05:21.403 & -31:49:19.96 & 6.615 & $2.22\pm1.00$ & $247\pm3$ & $420\pm5$ \\
J0305-O3E-PC-019 & 03:05:21.786 & -31:49:52.57 & 6.621 & $10.18\pm0.92$ & $280\pm33$ & $457\pm35$ \\
J0305-O3E-PC-020 & 03:05:21.820 & -31:50:25.36 & 6.617 & $2.64\pm1.00$ & $319\pm5$ & $582\pm10$ \\
J0305-O3E-PC-021 & 03:05:22.445 & -31:51:35.27 & 6.652 & $3.06\pm1.00$ & $357\pm6$ & $639\pm12$ \\
J0305-O3E-PC-022 & 03:05:21.222 & -31:49:59.74 & 6.617 & $4.94\pm0.77$ & $459\pm154$ & $720\pm169$ \\
J0305-O3E-PC-023 & 03:05:20.596 & -31:49:10.13 & 6.616 & $4.52\pm0.91$ & $375\pm51$ & $660\pm57$ \\
J0305-O3E-PC-024 & 03:05:21.041 & -31:49:58.66 & 6.618 & $6.06\pm0.75$ & $340\pm129$ & $544\pm139$ \\
J0305-O3E-PC-025 & 03:05:21.049 & -31:49:58.98 & 6.612 & $6.61\pm0.88$ & $258\pm40$ & $359\pm42$ \\
J0305-O3E-PC-026 & 03:05:20.424 & -31:49:11.65 & 6.612 & $3.79\pm0.80$ & $362\pm113$ & $605\pm123$ \\
J0305-O3E-PC-027 & 03:05:18.900 & -31:48:11.61 & 6.609 & $5.95\pm0.87$ & $604\pm101$ & $840\pm110$ \\
J0305-O3E-PC-028 & 03:05:19.997 & -31:50:19.68 & 6.619 & $1.91\pm0.70$ & $101\pm37$ & $123\pm37$ \\
J0305-O3E-PC-029 & 03:05:19.972 & -31:50:19.58 & 6.616 & $3.95\pm0.71$ & $183\pm62$ & $214\pm63$ \\
J0305-O3E-PC-030 & 03:05:19.955 & -31:50:19.26 & 6.624 & $2.44\pm0.84$ & $109\pm25$ & $157\pm25$ \\
J0305-O3E-PC-031 & 03:05:19.735 & -31:51:38.81 & 6.74 & $2.34\pm0.83$ & $319\pm82$ & $507\pm88$ \\
J0305-O3E-PC-032 & 03:05:18.612 & -31:50:10.65 & 6.54 & $3.89\pm1.00$ & $453\pm7$ & $836\pm14$ \\
J0305-O3E-PC-033 & 03:05:18.717 & -31:50:38.77 & 6.619 & $4.42\pm1.00$ & $204\pm2$ & $450\pm6$ \\
J0305-O3E-PC-034 & 03:05:18.729 & -31:50:38.53 & 6.616 & $3.99\pm0.71$ & $237\pm95$ & $339\pm98$ \\
J0305-O3E-PC-035 & 03:05:18.499 & -31:50:41.23 & 6.615 & $3.70\pm0.90$ & $326\pm62$ & $672\pm71$
\enddata

\end{deluxetable*}
\begin{deluxetable*}{ccccccc}
\tablehead{\colhead{Name} & \colhead{RA} & \colhead{Dec} & \colhead{$z$} & \colhead{log L$_{[OIII]}$} & \colhead{EW$_{[OIII]}$} & \colhead{EW$_{[OIII]+H\beta}$}\\ \colhead{} & \colhead{} & \colhead{} & \colhead{} & \colhead{10$^{42}$ erg s$^{-1}$} & \colhead{$\mathring{A}$} & \colhead{$\mathring{A}$}}
\tablecaption{Table \ref{table:o3pc} continued.}
\startdata
J0305-O3E-PC-036 & 03:05:16.291 & -31:49:32.29 & 6.599 & $2.42\pm0.76$ & $154\pm58$ & $253\pm60$ \\
J0305-O3E-PC-037 & 03:05:17.110 & -31:50:58.24 & 6.632 & $4.15\pm0.81$ & $314\pm82$ & $460\pm86$ \\
J0305-O3E-PC-038 & 03:05:16.530 & -31:50:22.65 & 6.667 & $10.79\pm0.85$ & $280\pm63$ & $462\pm67$ \\
J0305-O3E-PC-039 & 03:05:16.794 & -31:50:57.26 & 6.628 & $8.33\pm0.91$ & $513\pm70$ & $845\pm79$ \\
J0305-O3E-PC-040 & 03:05:15.698 & -31:50:37.57 & 6.616 & $3.29\pm0.88$ & $585\pm116$ & $1213\pm144$ \\
J0305-O3E-PC-041 & 03:05:16.824 & -31:53:27.01 & 6.503 & $6.85\pm0.89$ & $561\pm84$ & $852\pm92$ \\
J0305-O3E-PC-042 & 03:05:14.691 & -31:53:20.93 & 6.549 & $3.72\pm0.90$ & $435\pm71$ & $798\pm82$ \\
J0305-O3E-PC-043 & 03:05:13.138 & -31:52:52.45 & 6.67 & $3.25\pm0.76$ & $260\pm72$ & $303\pm73$ \\
J0305-O3E-PC-044 & 03:05:11.209 & -31:49:43.71 & 6.648 & $3.45\pm0.77$ & $308\pm94$ & $434\pm98$ \\
J0305-O3E-PC-045 & 03:05:10.353 & -31:49:04.97 & 6.815 & $4.47\pm0.82$ & $627\pm146$ & $834\pm157$ \\
J0305-O3E-PC-046 & 03:05:10.411 & -31:49:12.04 & 6.627 & $3.56\pm0.89$ & $408\pm65$ & $648\pm71$ \\
J0305-O3E-PC-047 & 03:05:12.580 & -31:52:52.76 & 6.673 & $7.38\pm1.00$ & $467\pm7$ & $718\pm12$ \\
J0305-O3E-PC-048 & 03:05:12.990 & -31:53:57.53 & 6.599 & $5.88\pm0.93$ & $441\pm52$ & $937\pm65$ \\
J0305-O3E-PC-049 & 03:05:10.165 & -31:49:11.53 & 6.755 & $3.68\pm0.78$ & $268\pm69$ & $316\pm70$ \\
J0305-O3E-PC-050 & 03:05:11.348 & -31:52:17.67 & 6.517 & $3.55\pm0.90$ & $1894\pm279$ & $3770\pm451$ \\
J0305-O3E-PC-051 & 03:05:10.246 & -31:52:00.90 & 6.596 & $5.99\pm1.00$ & $205\pm2$ & $306\pm3$ \\
J0305-O3E-PC-052 & 03:05:11.455 & -31:54:02.57 & 6.593 & $4.80\pm1.00$ & $523\pm7$ & $728\pm11$ \\
J0305-O3E-PC-053 & 03:05:07.985 & -31:49:11.78 & 6.748 & $3.16\pm1.00$ & $300\pm5$ & $492\pm8$
\enddata
\label{table:o3pc}
\end{deluxetable*}

\clearpage

\appendix
\section{Example Grism Spectrum}\label{sec:appA}
\begin{figure*}[h!]
\centering
\includegraphics[width=1.0\textwidth]{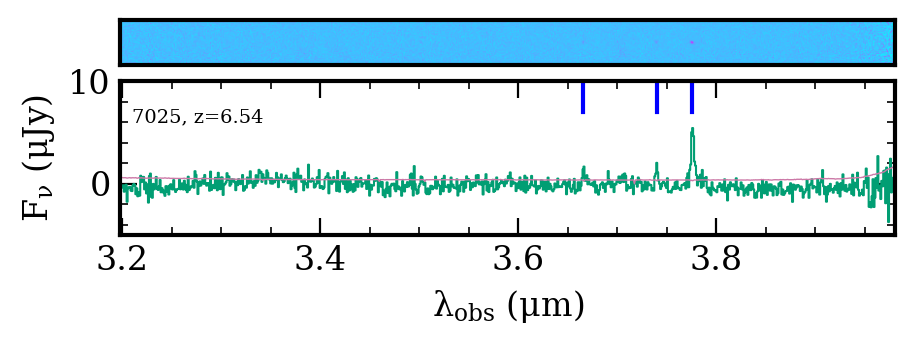}
\caption{\label{fig:appA} NIRCam/WFSS grism spectrum of source \#7025 at $z=6.54$. \textit{Top:} 2D continuum-subtracted spectrum of the source. \textit{Bottom:} The 1D optimally-extracted spectrum in green, with the error vector displayed in pink. Vertical blue lines indicate the locations of H$\beta$, \Oiii$\lambda$4959 and \Oiii$\lambda$5007, which are all securely detected. All galaxies in our sample have a secure detection of the \Oiii\, doublet.}
\end{figure*}

Here we display a representative example of one of our grism spectra, source \#7025 at $z=6.54$, demonstrating the quality of our high-confidence \Oiii\, emitters in Figure \ref{fig:appA}.

\section{Undetected LBGs}\label{sec:appB}
In \S\ref{sec:catalog} we discussed matching the LBG catalog from \citet{Champagne2023a} to the \Oiii\, catalog presented here, most of which are now either spectroscopically confirmed or eliminated as interlopers based on updated photometric redshifts.
However, three LBGs from \citet{Champagne2023a} are not detected in the F356W image so no spectrum was extracted for them.
\begin{figure*}
\centering
\includegraphics[width=1.0\textwidth]{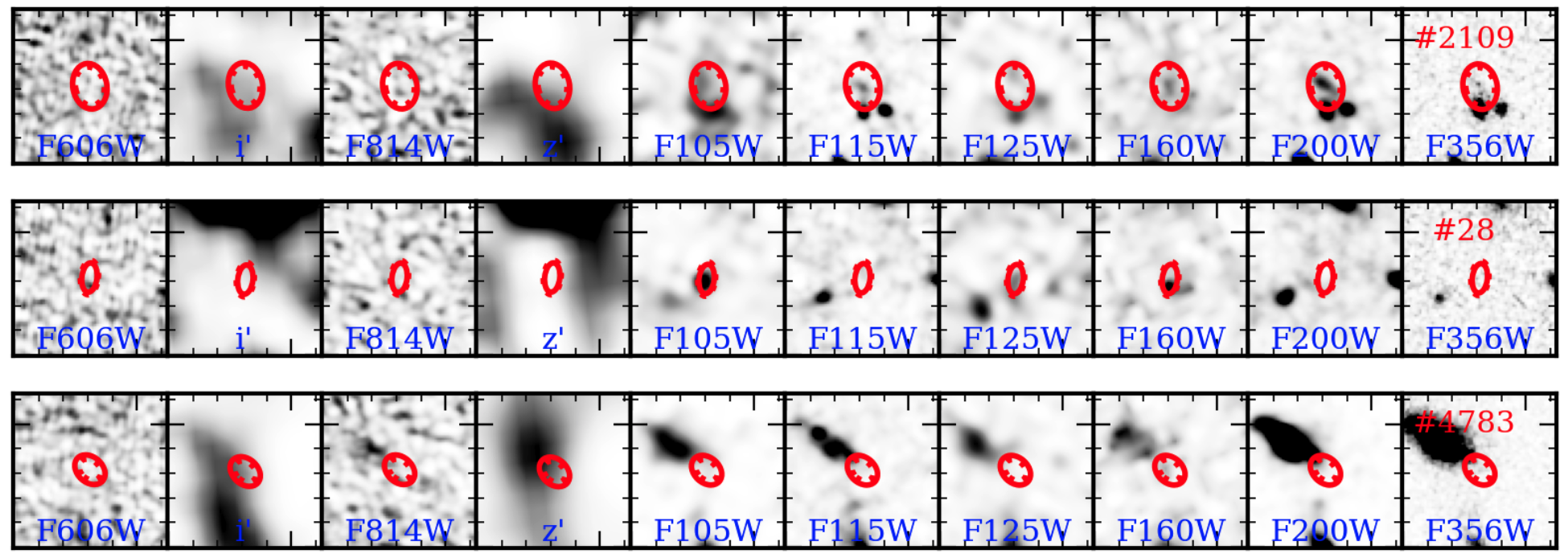}
\caption{\label{fig:appB} Cutouts in Subaru $i'$, $z$, \textit{HST}/ACS and WFC3 F606W, F814W, F105W, F125W, F160W, and \textit{JWST}/NIRCam F115W, F200W, F356W displayed in wavelength order. Cutouts are 2\arcsec$\times$2\arcsec, and the aperture used for photometry in the F160W detection image is displayed in red. All three sources shown are classified as ``marginal" LBGs by \citet{Champagne2023a} and are discarded in the present F356W sample.}
\end{figure*}
\citet{Champagne2023a} groups the \textit{HST}-detected LBGs into ``secure" and ``marginal" samples, where the ``secure" sample must satisfy: 1) the photometric redshift PDF $P(z)$ contains an 80\% probability of lying at $z>6$, 2) be detected in both F125W and F160W with $>5\sigma$ significance, and 3) displays a non-detection in F606W. 
For the ``marginal" case, this requirement is relaxed to $>40$\% of $P(z)$ contained at $z>6$ and a 3$\sigma$ detection in F125W and F160W. 
As can be seen in Figure \ref{fig:appB}, source \#2109 is likely not detected in F356W due to deblending from the nearby southern source.
Source \#28 is of low significance ($\sim4\sigma$) even in F160W, and does not appear visible in F356W.
Finally, source \#4783 is also probably picked up as an extended feature of the brighter northern source.
Since all three LBGs were in the ``marginal" sample, we disregard them as candidates in the F356W catalog.

\bibliography{main}




\suppressAffiliationsfalse
\allauthors
\end{document}